\documentstyle[11pt]{article}

\def\theequation{\arabic{section}.\arabic{equation}}

\renewcommand{\theequation}{\thesection.\arabic{equation}}
\def\baselinestretch{1.1}
\global\arraycolsep=1pt
\newcommand{\reg}{\mathop{\rm reg}\nolimits}

\newcounter{bean}
\newenvironment{romanlist}
    {
    \begin{list}{(\roman{bean})}
        {\usecounter{bean}
          \labelsep=1em
          \settowidth{\labelwidth}{(\thebean)}
          \addtolength{\labelwidth}{1.1ex} 
          \leftmargin=\labelwidth 
          \addtolength{\leftmargin}{\labelsep} }}
     {\end{list}}

\begin{document}

\input epsf

\hfill CPTH-S566.119, HUTP-97/A049, MIT-CTP-2689

\hfill hep-th/9711035

\hfill November, 1997

\vspace{20pt}

\begin{center}
{\large {\bf POSITIVITY CONSTRAINTS ON ANOMALIES
IN SUPERSYMMETRIC GAUGE THEORIES}}
\end{center}

\vspace{6pt}

\begin{center}
{\sl D. Anselmi}

{\it Centre de Physique Theorique, Ecole Polytechnique, F-91128 Palaiseau
Cedex, FRANCE}
\end{center}

\begin{center}
{\sl J. Erlich ~~and~~ D.Z. Freedman}

{\it Department of Mathematics and Center for Theoretical Physics,
Massachusetts Institute of Technology, Cambridge MA 02139, USA}
\end{center}


\begin{center}
{\sl A.A. Johansen}

{\it Lyman Laboratory, Harvard University, Cambridge, MA 02138, USA}
\end{center}

\vspace{8pt}

\begin{center}
{\bf Abstract}
\end{center}

\vspace{4pt}

The relation between the trace and $R$-current 
anomalies in supersymmetric theories implies that the
U$(1)_RF^2$, U$(1)_R$ and U$(1)_R^3$ anomalies which are
matched in studies of $N=1$ Seiberg duality satisfy positivity
constraints.  
Some constraints are rigorous and others
conjectured as four-dimensional generalizations of the
Zamolodchikov $c$-theorem.  
These constraints are tested in a
large number of $N=1$ supersymmetric 
gauge theories in the non-Abelian
Coulomb phase, and they are satisfied in all renormalizable
models with unique anomaly-free $R$-current, including those with
accidental symmetry.  
Most striking is the fact that the flow of
the Euler anomaly coefficient, $a_{UV}-a_{IR}$, is always
positive, as conjectured by Cardy.

\vfill\eject

\section{Introduction}

The computation of chiral anomalies of the $R$-current and
conserved flavor currents is one of the important tools used to
determine the non-perturbative infrared behavior of the many
supersymmetric gauge theories analyzed during the last few years.
The anomaly coefficients are subject to rigorous positivity
constraints by virtue of their relation to two-point functions of
currents and stress tensors, and to other constraints conjectured
in connection with possible four-dimensional analogues of the
Zamolodchikov c-theorem \cite{zamolo}. 
The two-point functions
have been considered \cite{gof97} as central functions whose 
ultraviolet
and infrared limits define central charges of super-conformal theories
at the endpoints of the renormalization group flow.  The
positivity conditions are reasonably well known from studies of
the trace anomaly for field theories in external backgrounds.  In
supersymmetric theories the trace anomaly of the stress tensor
and conservation anomaly of the $R$-current are closely related,
which leads \cite{bof97} to positivity constraints on chiral
anomalies.

Two studies of positivity constraints in the SU$(N_c)$ series of
SUSY gauge theories with $N_f$ fundamental quark flavors have
previously appeared.  
The first of these \cite{bastianelli}
analyzed the confined and free magnetic phases for $N_c < N_f <
3N_c/2$, while the basic techniques for computing the flow of
central charges when there is an interacting IR fixed point were
developed in \cite{bof97} and applied to the conformal phase for
$3N_c/2 < N_f < 3N_c.$ 
The most striking result of
\cite{bof97,bastianelli} was the positive flow,
$a_{UV}-a_{IR}>0$, of the coefficient $a(g(\mu))$ of the Euler
term in the trace anomaly in an external gravitational
background, where $g(\mu)$ is the gauge coupling at RG scale
$\mu$.  
This result agrees with the conjecture of Cardy
\cite{cardy} that the Euler anomaly obeys a $c$-theorem.
Positivity is also satisfied in all non-supersymmetric theories
tested \cite{cardy, Cappelli}.  
We shall refer to the inequality
$a_{UV} -a_{IR} >0$ as the $a$-theorem.  
The purpose of this
paper is to present an extensive exploration of the rigorous
positivity constraints and those associated with the $a$-theorem in
many supersymmetric gauge theories with interacting IR fixed
points (and some IR free models).  
We find that the $a$-theorem
and other constraints are satisfied in all renormalizable
theories we have examined, and there are other results of
interest.

In Sec. 2, which is largely a review of \cite{bof97}, the various
anomalies, the theoretical basis of the positivity constraints
and the computation of central charge flows are discussed.  
In Sec. 3 we discuss some general aspects of positivity 
constraints and the $a$-theorem
in models with $R$-charges uniquely 
fixed by classical conservation and cancellation of
internal anomalies.
In some models an accidental symmetry has
been postulated to preserve unitarity, and the central charges
must be corrected accordingly.
This is discussed in Sec. 4. 
In Sec. 5, the positivity constraints are tested in many examples of
renormalizable SUSY gauge models with uniquely determined 
$R$-charges.  
We also check the $a$-theorem for various types  
of flows between 
conformal fixed points.
The situation of some
non-renormalizable models is discussed in Sec.~6. 
There are
other models in which the conserved, anomaly free $R$-current is
not unique.  
Our methods are less precise in this case, but we 
discuss an example in Sec.~7.  
Sec. 8 contains a discussion of
results and conclusions.

\newcommand{\mysection}[1]{\setcounter{equation}{0}\section{#1}}
\renewcommand{\theequation}{\thesection.\arabic{equation}}
\newcommand{\PSbox}[3]{\mbox{\rule{0in}{#3}\includegraphics{#1}\hspace{#2}}}
\newcommand{\Tr}{\mathop{\rm Tr\,}\nolimits}

\newbox\circlebox
\setbox\circlebox=\hbox{\large$\bigcirc$}
\newdimen\circlewd
\newcommand{\circled}[1]{%
\setbox0=\hbox{$#1$}%
\circlewd=\wd0%
\box0\llap{\hbox to  
\circlewd{\hss\kern0.95pt\lower0pt\copy\circlebox\hss}}%
}

\section{Anomalies and Positivity Constraints}

The theoretical basis for the analysis of anomalies in supersymmetric
theories comes
from a combination of three fairly conventional ideas, namely

\renewcommand{\theenumi}{\Alph{enumi}}
\begin{enumerate}
\item The close relation between the trace anomaly of a 
four-dimensional field theory with external sources 
for flavor currents
  and stress tensor and the two point
  correlators $\langle J_{\mu} (x) J_{\nu}(y) \rangle$ and
  $\langle T_{\mu\nu} (x) T_{\rho \sigma} (y) \rangle$ and their
  central charges.

\item The close relation in a supersymmetric theory between the
  trace anomaly $\Theta = T^{\mu}_{\mu}$ and the anomalous
  divergence of the $R$-current $\partial_{\mu}R^{\mu}$.

\item The fact that anomalies of the $R$-current can be
  calculated at an infrared superconformal fixed point using 't
  Hooft anomaly matching. This is the standard procedure, and
  one way to explain it is to use the all orders anomaly free
  $S$-current of Kogan, Shifman, and Vainshtein \cite{kogan}.
\end{enumerate}
We now review these ideas briefly. More details are contained in
\cite{gof97,bof97}.

\subsection*{A. Trace anomaly and central charges}

We consider a supersymmetric gauge theory containing chiral
superfields $\Phi^{\alpha}_{i}$ in irreducible representations
$R_i$ of the gauge group $G$. 
To simplify the discussion we assume
that the superpotential $W=0$, but the treatment can be
generalized to include non-vanishing superpotential, and this
will be done in Sec. 2C below.

We consider a conserved current $J_{\mu}(x)$ for a non-anomalous
flavor symmetry $F$ of the theory, and we add a source
$B_{\mu}(x)$ for the current, effectively considering a new theory
with an additional gauged U$(1)$ symmetry but without kinetic
terms for $B_{\mu}$. The source can be introduced as an external
gauge superfield $B(x, \theta, \bar{\theta})$ so supersymmetry
is preserved. We also couple the theory to an external
supergravity background, contained in a superfield $H^{a} (x,
\theta, \bar{\theta})$, but we discuss only the vierbein $e^a_\mu(x)$ and the
component $V_{\mu} (x)$ which is the source for the $R^{\mu}$
current of the gauge theory.

The trace anomaly of the theory then contains several terms
\begin{equation}
\Theta=\frac{1}{2g^3}\tilde{\beta}(g)
(F_{\mu\nu}^a)^2 + \frac{1}{32 \pi^2} \tilde{b} (g) B^2_{\mu\nu}
+\frac{\tilde{c}(g)}{16\pi^2}(W_{\mu\nu\rho\sigma})^2-\frac{a(g)}
{16\pi^2}(\tilde{R}_{\mu\nu\rho\sigma})^2 +
\frac{\tilde{c}(g)}{6\pi^2}V_{\mu\nu}^2\, , \label{2.1}
\end{equation}
where $W_{\mu\nu\rho\sigma}$ is the Weyl tensor,
$\widetilde{R}_{\mu\nu\rho\sigma}$ is the dual of the curvature,
and $B_{\mu\nu}$ and $V_{\mu\nu}$ are the field strengths of
$B_{\mu}$ and $V_{\mu}$ respectively. All anomaly coefficients
depend on the coupling $g(\mu)$ at renormalization scale
$\mu$. The first term of (\ref{2.1}) is the internal trace
anomaly, where $\tilde{\beta}(g)$ is the numerator of the NSVZ beta
function \cite{nsvz}
\begin{equation}
\tilde{\beta}(g(\mu))=-\frac{g^3}{16\pi^2}\left[3T(G)-\sum_iT(R_i)
(1-\gamma_i(g(\mu)))\right] \, . \label{2.2}
\end{equation}
Here $T(G)$ and $T(R_i)$ are the Dynkin indices of the adjoint
representation of $G$ and the representation $R_i$ of the chiral
superfield $\Phi^{\alpha}_{i}$, and $\gamma_i/2$ is the anomalous
dimension of $\Phi^{\alpha}_{i}$.

The various external trace anomalies are contained in the three
coefficients $\tilde{b}(g)$, $\tilde{c}(g)$ and $a(g)$. 
The free field
({\em i.e.} one-loop) values of $\tilde{c}$ and $a$ have been known  
for many years
\cite{duff}. 
In a free theory of $N_0$ real scalars, $N_{1/2}$
Majorana spinors, and $N_1$ gauge vectors, the results are
\begin{eqnarray}
\label{2.3}
c&=&\frac{1}{120}(12 N_1 + 3N_{1/2} + N_0) \nonumber \\
a&=&\frac{1}{720}(124N_1+11N_{1/2}+2N_0)\,.
\end{eqnarray}
In a supersymmetric gauge theory with $N_v={\rm dim}\,G$ gauge multiplets 
and $N_\chi$ chiral
multiplets these values regroup as
\begin{equation} 
\label{2.4}
c_{UV}=\frac{1}{24}(3N_V+N_\chi)\qquad
a_{UV}=\frac{1}{48} (9N_V+N_\chi)\,.  
\end{equation}
If $T^j_i$ is the flavor matrix for the current $J_{\mu} (x)$
which is the $\bar{\theta} \theta$ component of the superfield
$\overline{\Phi}^i_{\alpha} T^j_i \Phi^{\alpha}_j$, and dim$R_i$
is the dimension of the representation $R_i$, 
the free-field value of $\tilde{b}$ is
\begin{equation}
  \label{2.5}
  b_{UV} =\sum_{i,j} (\dim R_i) T^j_i T^i_j
\end{equation}
The subscript $UV$ indicates that the free-field values are
reached in the ultraviolet limit of an asymptotically free
theory. 
Clearly $\tilde{c}$ and  $a$ count degrees of freedom of
the microscopic theory
with different weights for the various spin fields.

The current correlation function is
\begin{equation}
  \label{2.6}
  \langle J_{\mu} (x) J_{\nu} (0) \rangle = \frac{1}{16\pi^4}
  (\Box \delta_{\mu\nu} - \partial_{\mu}\partial_{\nu} )
  \frac{b(g(1/x))}{x^4}\,.
\end{equation}
It follows from reflection positivity or the Lehmann
representation as used in \cite{xavier} that the renormalization
group invariant central function \cite{bof97} $b(g(1/x))$ is
strictly positive. We assume that the theory in question has UV
and IR fixed points so that the following limits exist:
\begin{eqnarray}
\label{2.7}
b_{UV}&=&b(g_{UV})={\lim}_{x\rightarrow 0}\,b(g(1/x)) \nonumber \\
b_{IR}&=&b(g_{IR})={\lim}_{x\rightarrow\infty}\,b(g(1/x))\, . 
\end{eqnarray}
These endpoint values appear as central charges in the operator
product expansion of currents in the UV and IR
superconformal theories at the endpoints of the RG flow. 

The
correlator $\langle T_{\mu\nu} (x) T_{\rho\sigma}(0) \rangle$ has
the tensor decomposition \cite{gof97}
\begin{equation}
 \label{2.8}
\langle T_{\mu\nu}(x)T_{\rho\sigma}(0)\rangle=
\frac{1}{48\pi^4}\Pi_{\mu\nu\rho\sigma}\frac{c(g(1/x))}{x^4}
+\Pi_{\mu\nu}\Pi_{\rho\sigma}\frac{f(\log x\Lambda, g(1/x))}{x^4}\, ,
\end{equation}
where $\Pi_{\mu\nu} = ( \partial_{\mu} \partial_{\nu} -
\delta_{\mu\nu} \Box)$ and $\Pi_{\mu\nu\rho\sigma} = 2
\Pi_{\mu\nu} \Pi_{\rho\sigma} -3 (\Pi_{\mu\rho} \Pi_{\nu\sigma} +
\Pi_{\mu\sigma} \Pi_{\nu\rho})$ is the transverse traceless
projector and $\Lambda$ is the dynamical scale of the theory. 
The
central function $c(g(1/x))$ is a positive RG invariant
function. 
Its endpoint values $c_{UV}$ and $c_{IR}$ are also
central charges. 
The second tensor structure in (\ref{2.8}) arises
because of the internal trace anomaly. 
It is proportional to
$\widetilde{\beta} (g(1/x))$ and thus vanishes at critical points.

The important point is that there is a close relation between the
anomaly coefficients $\tilde{b} (g(\mu))$ and $\tilde{c} (g(\mu))$
and the central functions $b(g(\mu))$ and $c(g(\mu))$. 
Namely
$\tilde{b} (g(\mu))$ and $b(g(\mu))$ differ by terms proportional
to $\widetilde{\beta} (g(\mu))$, so they coincide at $RG$ fixed
points. 
The same holds for $\tilde{c} (g(\mu))$ and
$c(g(\mu))$. 
This means that the end-point values of the anomaly
coefficients are rigorously
positive. 
This is evident for the free field ultraviolet values in 
(\ref{2.3}-\ref{2.5}).  The infrared values $b_{IR}$ and $c_{IR}$ must also 
be positive, and this is an important check on the hypothesis that the 
long distance dynamics of a theory is governed by an interacting fixed point.

This important relation between trace anomaly coefficients and current
correlators was derived
in \cite{gof97,bof97} by an argument with the following ingredients:

\renewcommand{\theenumi}{\roman{enumi}}

\begin{enumerate}
\item Since the explicit scale derivative of a renormalized
  correlator corresponds to the insertion of the integrated
  trace anomaly, the $\langle J_{\mu} (x) J_{\nu} (0) \rangle $
  correlator satisfies
\begin{equation}
\label{2.9}
\mu\frac{\partial}{\partial\mu}\langle J_\mu(x)J_\nu(0)\rangle
=\frac{1}{8\pi^2}\tilde{b}(\mu)(\Box\delta_{\mu\nu}-
\partial_\mu\partial_\nu
)\delta^4(x)-\frac{\tilde{\beta}(g(\mu))}{2g^3}
\langle J_\mu(x)J_\nu(0)\int
    d^4z\,(F_{\rho\sigma}^a)^2\rangle \, .
\end{equation}

\item The central function $b(g(1/x))$ satisfies a standard
  homogeneous renormalization group equation, but $b(g(1/x))/x^4$
  requires additional regularization because it is singular at the
  origin. The regulated amplitude satisfies
\begin{equation}
\label{2.10}
\left.\mu\frac{\partial}{\partial\mu}\frac{b(g(1/x))}{x^4}\right|_{\reg}
=\frac{1}{8\pi^2}\hat{b}(g(\mu))\delta^4(x)+\left.
\frac{\beta(g(\mu))}{g^3}
\frac{b(g(1/x))}
{x^4}\right|_{{\reg}}\, . 
\end{equation}
where $\hat{b} (g(\mu))$ is associated with the overall
divergence at $x=0$.

\item Using the method of differential renormalization
  \cite{gof92} and the RG equation, one can resum a series in
  powers of $(\log x\mu)^k$ to derive a non-perturbative
  differential equation, namely
\begin{equation}
\label{2.11} 
\beta(g)\frac{\partial\hat{b}(g)}{\partial g}+2\hat{b}(g)=2b(g)\, .
\end{equation}
This shows that $\hat{b} (g(\mu))$ and the central function itself,
$b(g(\mu))$, coincide at fixed points. Comparing (\ref{2.9}) and
(\ref{2.10}) it is tempting to identify $\tilde{b} (g(\mu)) =
\hat{b} (g(\mu))$, but this also holds only at fixed points since
we cannot exclude possible local $\delta^4(x)$ terms in the
$\langle JJ \int F^2 \rangle$ correlator. It is easy to see that
contributions to $\langle JJ \int F^2 \rangle$ begin at order
$g(\mu)^4$. It is assumed that the local terms have no
singularities which could cancel the zero of $\widetilde{\beta}
(g)$.
\end{enumerate}

The anomaly coefficient $a(g(\mu))$ is related to 3-point
correlators of the stress tensor \cite{Osban} rather than to
$\langle T_{\mu\nu} T_{\rho\sigma} \rangle$. However it is clear
that $a(g(\mu))$ is significant, and that the
fixed point values $a_{UV}$, $b_{UV}$, $c_{UV}$ and $a_{IR}$,
$b_{IR}$, $c_{IR}$ are important quantities which characterize
the superconformal theories at the fixed points of the RG flow.

{\bf $c$-theorems}:\hspace*{1em} In two dimensions Zamolodchikov
established the $c$-theorem by constructing a function
$C(g(\mu))$ as a linear combination of (suitably scaled) $\langle
T_{zz} T_{zz}\rangle$, $\langle T_{zz} \Theta\rangle$ and $\langle
\Theta\Theta \rangle$ correlators which satisfies:
\begin{eqnarray}
\label{2.12}
\mu \frac{\partial}{\!\!\! \partial \mu} C(g(\mu)) &\geq& 0 \nonumber \\
\frac{\partial}{ \partial g}C(g(\mu))\left|_{g = g^{\ast}}
                                                \right.&=&0\\
C(g^{\ast}) &=& c^{\ast} \nonumber
\end{eqnarray}
where $c^{\ast}$ is the Virasoro central charge of the critical theory
at the fixed point $g=g^{\ast}$ or, equivalently, the fixed point
value of the external trace anomaly coefficient
\begin{equation}
\label{2.125}
  \Theta = \frac{1}{24 \pi} c^* R\,,
\end{equation}
where $R$ is the scalar curvature.
The properties (\ref{2.12}) imply $c_{UV} - c_{IR}>0$ which is the form in
which the $c$-theorem is usually tested \cite{ctest}. 
The
ingredients of Zamolodchikov's proof of these properties are
conservation Ward identities, rotational symmetry, reflection
positivity, and Wilsonian renormalizability. 
There is a similar
proof \cite{xavier} of a $k$-theorem for the central charges of
conserved currents, which leads to $b_{UV} - b_{IR} \geq 0$ in
our notation.  
There are alternative proofs \cite{Cappelli,xavier} of
the $c$ and $k$-theorems in two dimensions based on the Lehmann
representation for the invariant amplitudes in the decomposition
of $<T_{\mu\nu}(p)T_{\rho\sigma}(-p)>$ and $<J_\mu(p)J_\nu(-p)>$.

The techniques used in the two-dimensional case cannot be extended
to four dimensions \cite{cardy,Cappelli}, 
and it has not so far been possible to
construct any $C$-function for four-dimensional theories which
satisfies (\ref{2.11}). 
The best thing one now has is Cardy's
conjecture \cite{cardy} that there is a universal $c$-theorem
based on the Euler anomaly, so that $a_{UV} - a_{IR} >0$ in all
theories. 
There is theoretical support for this conjecture \cite{Osban}, 
and empirical support by direct test in models where the
infrared dynamics is understood. The $a$-theorem is true in all
models so far tested which include
\begin{romanlist}
\item SU$(N_c)$ QCD with $N^2_c -1$
  gluons and $N_fN_c$ quarks \cite{cardy}. An infrared realization
  as a confined theory with chiral symmetry breaking and $N^2_f -1$
  decoupled Goldstone bosons is assumed.

\item QCD at large $N_c$ with $N_f = 11 N_{c}/2 -k$ near the
  asymptotic freedom limit. The infrared limit
  is computable in perturbation theory because of the well known
  close two-loop fixed point \cite{Banks-Zaks}. Actually $a_{UV} - a_{IR} =0$
  to order $1/N^2_c$ for reasons we discuss below.

\item SU$(N_c)\; N=1$ SUSY QCD in the confined and free magnetic phase for
  $N_c\leq N_f \leq  \frac{3N_c}{2}$ \cite{bastianelli}.

\item SU$(N_c)\; N=1$ SUSY QCD in the non-Abelian Coulomb phase for
  $\frac{3 N_c}{2} < N_f < 3 N_c$ \cite{bof97}.

\end{romanlist}

One may take a more general empirical approach and test whether
other $c$-theorem candidates such as the total flow $b_{UV} -
b_{IR}$ and $c_{UV} - c_{IR}$ (or possible linear combinations
with $a_{UV} -a_{IR}$) are positive in the 
models above.
It is known that $c_{UV}-c_{IR}$ is positive in the
situations i)
\cite{Cappelli} and iii) \cite{bastianelli} above, but negative in situation
ii) \cite{Cappelli} and changes sign from positive to negative as $N_f$
increases in the theories of iv). Thus a universal
``$c$-theorem'' is ruled out. In the Appendix below we present
brief calculations to show that a $b$-theorem cannot hold in 
situations i)--iii) above, and it is known \cite{bof97} not to
hold in situation iv).

Thus the $a$-theorem, $a_{UV}-a_{IR} >0$, emerges as the only
surviving candidate for a universal theorem in four
dimensions. The desired physical interpretation requires the
existence of an $A$-function $A(g(\mu))$ which decreases monotonically from
$a_{UV}$ to $a_{IR}$ and counts effective degrees of freedom at a
given scale. Thus the relation $a_{UV}-a_{IR} >0$ would make
little physical sense unless $a_{IR}$ is positive. Indeed it has
been argued \cite{l-a} that $a(g(\mu))$ is positive at critical
points if a conjectured quantum extension of the weak energy
condition of general relativity is valid. 

Let us now summarize this discussion of the positivity properties
of trace anomaly coefficients. The free-field values $a_{UV}$,
$b_{UV}$, $c_{UV}$ are automatically positive. Positivity is
rigorously required for $b_{IR}$ and $c_{IR}$, and  it is a
useful test of our understanding of the infrared dynamics to check
this property in models. We will also explore the conjectured
$a$-theorem and the related condition $a_{IR} >0$. We will also show
that the ``data''
for $N=1$ SUSY gauge theories in the non-Abelian Coulomb phase 
imply that there is no linear combination $u(a_{UV}-a_{IR}) +
v(c_{UV}- c_{IR})$ which is positive in all models
(except for $v=0, ~u>0$). 

\subsection*{B. Relation between 
$\Theta$ and $\partial_{\mu}R^{\mu}$
  anomalies in SUSY/SG.}

In a supersymmetric theory in the external U$(1)$ gauge and
supergravity backgrounds discussed above, the divergence of the
$R^{\mu}$ current and the trace of the stress
tensor are components of a single superfield. Therefore the
supersymmetry partner of the trace anomaly $\Theta$ of
(\ref{2.1}) is
\begin{equation}
  \label{2.13}
  \partial_{\mu} (\sqrt{g} R^{\mu}) = - \frac{1}{3g^3}
  \widetilde{\beta} (g) (F \widetilde{F}) -
  \frac{\tilde{b}(g)}{48\pi^2} (B \widetilde{B}) +
  \frac{\tilde{c} (g) - a(g)}{24 \pi^2} R \widetilde{R} +
  \frac{5a(g) - 3 \tilde{c} (g)}{9 \pi^2} (V \widetilde{V})
\end{equation}
where $R$ and $\widetilde{R}$ on the right hand side are the
curvature tensor and its dual. The ratio $-2/3$ between the
first two terms of (\ref{2.1}) and (\ref{2.13}) is well known in
global supersymmetry, but the detailed relation of the anomaly
coefficients of the gravitational section was first derived in
\cite{bof97} by evaluating the appropriate components of the curved
superspace anomaly equation
\begin{equation}
  \label{2.14}
  \bar{D}^{\dot\alpha} J_{\alpha \dot{\alpha}} =
  \frac{1}{24\pi^2} 
\left(\tilde{c} W^2 - a \Xi \right)
\end{equation}
where $J_{\alpha \dot{\alpha}}$, $W^2$ and $\Xi$ are the
supercurrent, super-Weyl, and super-Euler superfields
respectively. This equation shows that all gravitational
anomalies are described by the two functions $\tilde{c}(g)$ and
$a(g)$, and this is also the reason why the coefficients of the third
and fifth terms of (\ref{2.1}) are related. An alternate
derivation of (\ref{2.13}) which does not require superspace
technology was also given in \cite{bof97}.

The last three terms of (\ref{2.13}) are essentially the same as
the anomalies usually computed in studies of $N=1$ Seiberg duality. It
is this fact that leads to immediate positivity constraints on
supersymmetry anomalies which we can test easily in the various
models in the literature which flow to infrared fixed points.

\subsection*{C. Computing infrared anomaly coefficients.}

In this section we discuss how the infrared central charges
 $b_{IR}$, $c_{IR}$ and $a_{IR}$ are related to the
 conventional U$(1)_R F^2$, U$(1)_R$ and U$(1)^3_R$
 anomalies. This is already quite clear, and some readers may
 wish to jump ahead to the final formulae at the end. However we
 do think that it is useful to derive this relation using the
 formalism of the all-orders anomaly-free $S^{\mu}$ current
 introduced in \cite{kogan}. The external
 anomalies of this current can be clearly seen to agree in the
 infrared limit with those of the $R^{\mu}$ current which is in
 the same multiplet as the stress tensor, and thus part of the
 $N=1$ superconformal algebra of the infrared fixed point theory.
A very clear
explanation of the $S^{\mu}$ current is given in
Section~3 of \cite{kogan} for the case of general gauge
group $G$ and arbitrary superpotential $W( \phi)$.  We summarize
and exemplify the argument for the slightly simpler case of cubic
$W (\phi)$.  

Gaugino fields are denoted by $\lambda^a (x) \, , \, a=1, \ldots
,\dim G$, and scalar and fermionic components of
$\Phi^{\alpha}_i(x)$ by $\phi^{\alpha}_i (x) $ and
$\psi^{\alpha}_{i} (x)$ respectively. The canonical $R^{\mu}$
current (which is the partner of the stress tensor), and the matter
Konishi currents $K^{\mu}_i$ for each representation are
\begin{eqnarray}
\label{2.15}
  R^{\mu} &=& \frac{1}{2} \overline{\lambda}^a \gamma^{\mu}
               \gamma^5 \lambda^a - \frac{1}{6} \sum_i
               \overline{\psi}^i_{\alpha} \gamma^{\mu} 
               \gamma^5 \psi^{\alpha}_{i} + \frac{2}{3}
               \sum_i \overline{\phi}^i_{\alpha} 
               \stackrel{\leftrightarrow}{D}_{\mu}
               \phi^{\alpha}_i \nonumber\\[1ex]
  K^{\mu}_i &=& \frac{1}{2} \sum_i \overline{\psi}_{\alpha}^i \gamma^{\mu}
                \gamma^5 \psi^{\alpha}_i +
                \sum_i \overline{\phi}^i_{\alpha}
                \stackrel{\leftrightarrow}{D}_{\mu} \phi^{\alpha}_i \, .
\end{eqnarray}
Conservation of the Konishi current is spoiled by a classical
violation for any non-vanishing $W$ and a $1$-loop exact chiral
anomaly.  The internal anomaly of $R^{\mu}$ in (\ref{2.13}) can
also be generalized to include $W$.  The divergences of these
currents are then (external sources are dropped)
\begin{eqnarray}
\label{2.16}
  \partial_{\mu} K^{\mu}_i &=& \Phi^{\alpha}_i 
       \frac{\partial W}{\partial \Phi^{\alpha}_i} \, \Bigg|
        \, + \frac{T(R_i)}{16 \pi^2} F
        \widetilde{F}\\[1ex]
   \partial_{\mu} R^{\mu} &=& \frac{1}{3} \sum_i \gamma_i 
         \Phi^{\alpha}_i
         \frac{\partial W}{\partial \Phi^{\alpha}_i} \, \Bigg|
         + \frac{1}{48 \pi^2} \left[ 3T(G) - \sum_i T(R_i) (1- \gamma_i)
         \right] F \widetilde{F} \label{2.17}
\end{eqnarray}
where $\mid$ indicates the $\theta^2$ component of the
superfield minus its adjoint. 
 The anomaly-free $R$
current usually stated in the literature for any given model is a
specific linear combination (assumed unique here)
\begin{equation}
\label{2.18}
  S^{\mu}_0 = R^{\mu} + \frac{1}{3} \sum_i \gamma_i^{\ast} K^{\mu}_i\,.
\end{equation}
which is conserved classically and non-anomalous to one-loop
order. This means that all terms in its divergence,
\begin{equation}
\label{2.19}
  \partial_{\mu} S^{\mu}_0 = \frac{1}{3} \sum_i \left(
    \gamma^{\ast}_i + \gamma_i \right) \Phi^{\alpha}_i
   \frac{\partial W}{\partial \Phi^{\alpha}_i} \Bigg|
   + \frac{1}{48 \pi^2} \left[ 3T (G) - \sum_i T(R_i)
     (1-(\gamma^{\ast}_i + \gamma_i)) \right] F^a_{\mu v} 
    \widetilde{F}^{\mu va} \, ,
\end{equation}
cancel except those with coefficients $\gamma_i$. There
is then a unique (flavor singlet) all-order conserved current
\begin{equation}
\label{2.20}
  S^{\mu} = R^{\mu} + \frac{1}{3} \sum_i (\gamma^{\ast}_i - \gamma_i) 
  K^{\mu}_i
 \end{equation}
Its divergence vanishes,
\begin{equation}
\label{2.21}
  \partial_{\mu} S^{\mu} = \frac{1}{3} \sum_i \gamma^{\ast}_{i}
  \phi^{\alpha}_i \frac{\partial W}{\partial \phi^{\alpha}_i}
    \bigg| + \frac{1}{48 \pi^2}\left[3T (G) - \sum_i T(R_i) (1
    -\gamma^{\ast}_i ) \right] F
    \widetilde{F} =0\,,
\end{equation}
and the vanishing of the coefficients of $F \widetilde{F}$ and
the independent cubic terms means that the $\gamma^{\ast}_i$ are
the unique set of numbers which make the gauge and various Yukawa
beta functions vanish.  
The
$\gamma^{\ast}_i$ then have the physical interpretation as IR
anomalous dimensions of the superfields $\phi^{\alpha}_i$,
assuming that there is an IR fixed point. 
In the infrared
limit, $\gamma_i \to \gamma^{\ast}_i$ in (\ref{2.20}), and $S^{\mu} \to
R^{\mu}$. 
It is worth noting that the coefficient in front of the 
Konishi current in (\ref{2.20}) is a manifestation 
of positive anomalous dimension of the anomalous Konishi 
current \cite{univer}.
In physical correlators the infrared limit can be associated 
with large distance behavior.  
Therefore in the infrared (large distance) limit 
of correlators with an insertion of $R_{\mu}=
S_{\mu}-\frac{1}{3} \sum_i (\gamma^{\ast}_i - \gamma_i) 
K^{\mu}_i$ 
the contribution of the Konishi current decreases faster
than the contribution of the $S_{\mu}$ current
which has no anomalous dimension.
Thus the $S^{\mu}$ and $R^{\mu}$ operators and their
correlators agree in the long distance limit, as is required at
the superconformal $IR$ fixed point.
In the free $UV$ limit
$\gamma_i \to 0$, and $S^{\mu} \to S^{\mu}_0$. As we will see
shortly this means that external anomalies of $S^{\mu}$ coincide
with those computed in the literature.

We distinguish three classes of models in which one obtains unique
$S^{\mu}_0$ and $S^{\mu}$ currents.  The first is the set of
models with chiral fields in $N_f$ copies of a single (real)
irreducible representation $R$ (or $N_f$ fields in 
$R\oplus\overline{R})$ and no superpotential.  It is
easy to see that the unique $S^{\mu}$ current in these two cases
is
\begin{eqnarray}
\label{2.22}
  S^{\mu} &=& R^{\mu} + \frac{1}{3} \left( 1-
              \frac{3T(G)}{N_f T(R)} - \gamma( g(\mu) )\right)
            \sum_i K^{\mu}_i  \\
  S^{\mu} &=& R^{\mu} + \frac{1}{3} \left( 1-
              \frac{3T(G)}{2N_f T(R)} - \gamma  (g (\mu))
            \right) \sum_i (K^{\mu}_i + \widetilde{K}^{\mu}_i )\nonumber
\end{eqnarray}
(where $K^{\mu}_i$ and $\widetilde{K}^{\mu}_i$ are the Konishi
currents of fields in the $R$ and $\overline{R}$ representations,
respectively, and we use $T(R) = T(\overline{R})$ and $\gamma =
\widetilde{\gamma}$).  
Comparing with (\ref{2.2}), one can see
that the coefficient of the Konishi terms is proportional to
$\widetilde{\beta}(g (\mu ))$ and thus vanishes in the infrared limit
if there is a fixed point.

The second class of models are those of Kutasov \cite{kut}
and generalizations \cite{intr,leigh} in which we add a
superfield $X$ in the adjoint representation to the previous
matter content and take $W= f \Tr X^3$.  We let $K^{\mu}_X$ and
$\gamma_X$ denote the Konishi current and anomalous dimension for the
adjoint fields.  The procedure outlined above leads to the unique
currents
\begin{eqnarray}
\label{Scurr}
    S^{\mu} &=& R^{\mu} + \frac{1}{3} \left( 1-
              \frac{2T(G)}{N_f T(R)} - \gamma (g,f) \right)
            \sum_i K^{\mu}_i - \frac{1}{3} \gamma_X
            K^{\mu}_X \nonumber \\
  S^{\mu} &=& R^{\mu} + \frac{1}{3} \left( 1-
              \frac{T(G)}{N_f T(R)} - \gamma (g,f)\right)
           \sum_i (K^{\mu}_i + \widetilde{K}^{\mu}_i ) -
           \frac{1}{3} \gamma_X K^{\mu}_X 
\end{eqnarray}
for the cases of representations $R \oplus adj$, and $R\oplus
\overline{R} \oplus adj$, respectively.  
If there is an $IR$ fixed
point, then both $\beta_f = 3f \gamma_X/2$ and $\tilde{\beta}(g)$ given
in (\ref{2.2}) must vanish, and it is easy to see that all
coefficients of the Konishi terms in (\ref{Scurr}) vanish if this
occurs.  
The procedure may be extended to more general models
with $W=f \Tr X^{k+1}, \quad k>2$, using the modification of
(\ref{2.19}) (see Section 3 of \cite{kogan}) for general
superpotentials.

Another common class of models resembles the ``magnetic'' version
of SU$(N_c)$ SUSY QCD. There are $N_f$ flavors of quark and
anti-quark fields $q$ and $\tilde{q}$ in conjugate
representations $R^{\prime}$ and $\overline{R}^{\prime}$ of a dual
gauge group $G^{\prime}$ plus a gauge singlet $M$ in the $(N_f,
\bar{N}_f)$ representation of the flavor group. The models have a
cubic superpotential $W= f \tilde{q} M q$. 
In this case the
unique $S^{\mu}$ current is
\begin{equation}
 \label{SScur}
 S^{\mu} = R^{\mu} + \frac{1}{3} \left( 1 - \frac{3T (G')}{2N_f
 T(R')} - \gamma_q \right) ( K^{\mu}_i + \widetilde{K}^{\mu}_i
-2K^{\mu}_M ) - \frac{1}{3} (2\gamma_q + \gamma_M) K^{\mu}_M,
\end{equation}
and one can check again that the coefficients of independent Konishi
currents vanish exactly when $\beta_g = \beta_f = 0$.

Because the operator $S^{\mu}$ is exactly conserved without
internal anomalies, 't Hooft anomaly matching \cite{hooft} can be
applied to calculate the anomalies of its matrix elements with
other exactly conserved currents, such as $\partial_\mu
\langle S^\mu
T^{\rho\sigma} T^{\lambda\tau}\rangle $. 
One argument for this (Sec 3
of \cite{bof97}) is the following. The operator
equation $\partial_\mu S^\mu =
0$ holds in the absence of sources, and it must remain local 
when sources are introduced. 
For an
external metric source dimensional and symmetry considerations
restrict the possible form of the matrix element to
\begin{equation}
\label{2.25}
  \langle \partial_\mu S^\mu(x)\rangle = s_0 R\tilde{R}(x) 
\end{equation}
where the right hand side is local. 
{\em A priori} $s_0(g(\mu))$ could
depend on the RG scale $\mu$.  However, $S^\mu$ in this case is an
RG invariant operator, so matrix elements cannot depend on
$g(\mu)$. Therefore $s_0$ must be a constant, hence 1-loop exact.
If we now use the fact that $S$ and $R$ coincide at long
distances we have the chain of equalities
\begin{equation}
\label{2.26}
  \partial\langle RTT \rangle_{IR} = \partial\langle STT
  \rangle_{IR} = \partial\langle STT \rangle_{UV} = \partial\langle
  S_0 TT \rangle
\end{equation}
where the last term simply includes the one loop graphs of the
current $S_0$ and gives the U(1)$_R$ anomaly coefficient quoted
in the literature. Similar arguments
justify the conventional calculation of of U$(1)_RFF$ and
U$(1)_R^3$ anomalies.

{\bf Formulae for anomaly coefficients:}
The previous discussion enables us to write simple formulae for the infrared
values of the anomaly coefficients in terms of the anomaly-free $R$-charges
quoted in the literature.  
For a chiral superfield $\Phi_i^\alpha$ in the
representation $R_i$ of dimension ${\rm dim}\,R_i$ the $R$-charge $r_i$ is
related to $\gamma_i^\ast$ in the $S_0^\mu$ current (\ref{2.18}) by
$r_i=(2+\gamma_i^\ast)/3$.

The quantities $b_{IR}$, $c_{IR}$ and $a_{IR}$ are the infrared values
of the trace anomaly coefficients $\tilde{b}$, $\tilde{c}$ and
$a$ in (\ref{2.1}).  
They are normalized by the free field
values in (\ref{2.4}) and (\ref{2.5}) and are related to $R$-current
anomalies by (\ref{2.13}).  
One then obtains

\begin{eqnarray}
  b_{IR}&=&-3U(1)_RF^2=3\sum_{ij}({\rm
             dim}\,R_i)(1-r_i) T_i^j\,T_j^i \label{abcIR} \nonumber \\
  c_{IR}-a_{IR}&=&-\frac{1}{16} U(1)_R=-\frac{1}{16}({\rm
             dim}\,G+ \sum_i({\rm dim}\,R_i)(r_i-1)) \nonumber \\
  5a_{IR}-3c_{IR}&=&\frac{9}{16}U(1)_R^3=\frac{9}{16}({\rm
             dim}\,G+\sum_i ({\rm dim}\,R_i)(r_i-1)^3)  \\
  c_{IR}&=&\frac{1}{32}(9U(1)_R^3-5U(1)_R)=\frac{1}{32}(4{\rm
              dim}\,G+\sum_i ({\rm dim}\,R_i)(1-r_i)(5-9(1-r_i)^2)
              \nonumber \\
  a_{IR}&=&\frac{3}{32}(3U(1)_R^3-U(1)_R)=\frac{3}{32}(2{\rm
           dim}\,G+\sum_i ({\rm dim}\,R_i)(1-r_i)(1-3(1-r_i)^2)).
\nonumber
\end{eqnarray}

Note that the $R$-charge of the fermionic component of $\Phi_i^\alpha$ is
$r_i-1$ and appears in these formulae, which are valid for theories in an
interacting conformal phase with unique anomaly free $R$-charges and no
accidental symmetry.  The treatment is extended to include accidental
symmetry and theories with nonunique $R$-charge in later sections.

The hypothesis that there is a nontrivial infrared fixed point in
any given model is established by several consistency tests which in the
past did not include the positivity conditions we have discussed.  The
set of infrared $R$-charges assigned in the literature is not guaranteed
to produce positive $b_{IR}$, $c_{IR}$, $a_{IR}$ so the positivity
constraints provide an additional check that the hypothesis of an
interacting fixed point is correct.

The corresponding UV quantities are obtained from (\ref{abcIR}) by replacing
$r_i\rightarrow 2/3$, and one can check that (\ref{2.4}) and (\ref{2.5}) are 
reproduced
when this is done.  Thus for flows without gauge symmetry breaking the total 
flow of the central charges
from the UV to the IR is due to the difference between the
canonical and non-anomalous $R$-charges, and are given by the following
formulae:
\begin{eqnarray}
b_{UV}-b_{IR}&=&3\sum_{ij}({\rm dim}\,R_i)[(r_i-\frac{2}{3})
          T_i^jT_j^i] \label{bbflow} \\
c_{UV}-c_{IR}&=&\frac{1}{384}\sum_i({\rm dim}\,R_i)(2-3r_i) [
         (7-6r_i)^2 -17] \label{ccflow} \\
a_{UV}-a_{IR}&=&\frac{1}{96}\sum_i({\rm dim}\,R_i)(3r_i-2)^2(5-3r_i)
\label{aaflow}.
\end{eqnarray}
Higgs flows with spontaneous symmetry breaking of gauge symmetry are
studied in Section 3.

There is a rather interesting aspect of the formulae (\ref{bbflow}), 
(\ref{ccflow}), (\ref{aaflow}) for
central charge flows.
In perturbation theory about a UV free
fixed point the quantity ($2-3r_i$) is of order $g^2$.
Thus our
formulas are consistent with the 2-loop calculations
of \cite{jack}
who found that radiative corrections to $c(g)$ begin at 2-loop
order (and quantitatively agree \cite{bof97} with the perturbative
limit of (\ref{ccflow})), while corrections to $a(g)$ vanish at 2-loop
order.
The ``input'' to 
 (\ref{aaflow}) comes from 1-loop chiral anomalies, so
it is curious that the formula for $a_{UV}-a_{IR}$ ``knows'' about 2-loop 
curved space computations.

The perturbative structure becomes more significant when we
consider the physical requirement that a $c$-function must be
stationary at a fixed point, and that Zamolodchikov's
$C$-function actually satisfies $\frac{\partial}{\partial g}
C(g) = 0$ at a fixed point. 
A monotonic interpolating $A$-function is not known in four dimensions
but one can consider a candidate
$A$-function obtained from $a_{IR}$
in (\ref{abcIR}) by replacing the infrared values of $r_i$ by their values
calculated along the flow, {\em i.e.} $r_i \to (2 + \gamma_i
(g(\mu)))/3$. 
This candidate $A$-function naturally satisfies
Zamolodchikov's stationarity condition at weak coupling. The
analogous candidate $C$-function obtained from $c_{IR}$ of (\ref{abcIR})
does not.

\def\theequation{\arabic{section}.\arabic{equation}}
\renewcommand{\theequation}{\thesection.\arabic{equation}}
\def\baselinestretch{1.1}
\global\arraycolsep=1pt
\oddsidemargin .20in
\evensidemargin .5in
\topmargin 0in
\textwidth 6.25in
\textheight 8.5in

\renewcommand{\mag}{\mathop {\rm mag}}

\newcounter{been}
\newenvironment{alphlist}
    {
    \begin{list}{\alph{been}.)}
        {\usecounter{been}
          \labelsep=1em
          \settowidth{\labelwidth}{(\thebean)}
          \addtolength{\labelwidth}{1.1ex} 
          \leftmargin=\labelwidth 
          \addtolength{\leftmargin}{\labelsep} }}
     {\end{list}}

\newcommand{\drawsquare}[2]{\hbox{%
\rule{#2pt}{#1pt}\hskip-#2pt
\rule{#1pt}{#2pt}\hskip-#1pt
\rule[#1pt]{#1pt}{#2pt}}\rule[#1pt]{#2pt}{#2pt}\hskip-#2pt
\rule{#2pt}{#1pt}}

\newcommand{\Yfund}{\raisebox{-.5pt}{\drawsquare{6.5}{0.4}}}
\newcommand{\Ysymm}{\raisebox{-.5pt}{\drawsquare{6.5}{0.4}}\hskip-0.4pt%
        \raisebox{-.5pt}{\drawsquare{6.5}{0.4}}}
\newcommand{\Yasymm}{\raisebox{-3.5pt}{\drawsquare{6.5}{0.4}}\hskip-6.9pt%
        \raisebox{3pt}{\drawsquare{6.5}{0.4}}}

\section{Models with Unique $R$-charge}

In this section we discuss
the positivity conditions $b_{IR} >0$,
$c_{IR}>0$, $a_{IR}>0$ and $a_{UV} - a_{IR} >0$
in a large set of
models in the literature where the anomaly-free
$R$-charge is unique.
While some of these models will be considered in more detail
in the next two sections, here we are going to analyze
some general aspects.
It is worth emphasizing that even though the positivity of
$b_{IR}$ and $c_{IR}$ follows generally
from unitarity constraints, the fact that they turn out to be positive
in our approach is additional evidence that
our understanding of the infrared dynamics is correct.

The positivity constraint $a_{UV} - a_{IR}>0$
deserves some comments.
As explained above, the gravitational effective action depends on
the functions $a$ and $c$.
It is natural to assume that a candidate $C$-function
measuring the irreversibility of the RG flow
may be a universal model independent linear combination
$C=ua+vc$.
We are going to show that the only combination
$C=ua+vc$
which satisfies
$\Delta
C= u(a_{UV} -a_{IR}) + v(c_{UV} -c_{IR})\geq 0$ for all models
is just $C=a$.
First note that
since there are theories (e.g.
SU$(N_c)$ SUSY QCD with $N_f < 3 N_c$) with $c_{UV} -
c_{IR}$ of either sign \cite{bof97}
and $a_{UV}-a_{IR}$ positive, one must take $u>0$.
It is then sufficient to assume $u=1$.
Consider the electric version of Seiberg's
SU$(N_c)$ QCD with $N_f$ fundamental
flavors in the conformal window, $3N_c/2<N_f<3N_c$.
In
the weak
coupling limit $N_c$, $N_f \to \infty$ and $N_c/N_f \to 3$, the
work of \cite{bof97} shows that $\Delta c < 0$
and $0\leq
\Delta a << \mid \Delta c \mid$.
So we have $v\leq 0$.
On the other hand
in the weak coupling limit $N_f \to \infty$ and $N_c/N_f \to 3/2$ of
the magnetic theory one can see that
$0\leq
\Delta a <<\Delta c$  so we have $v\geq 0$.
Then $v=0$, and
$a_{UV} - a_{IR} >0$ is the only universal $a$-theorem candidate.

Below we state simple sufficient conditions for the
positivity
constraints $b_{IR} >0$,
$c_{IR}>0$, $a_{IR}>0$, and also for
$a_{UV} - a_{IR}>0$ in the case of RG flows from a
free ultraviolet to an infrared fixed point.
Remarkably enough, 
these sufficient conditions can be quickly seen to be satisfied 
in most of the
conformal window of all renormalizable theories
that we have analyzed.  
Closer examination is required for cases with accidental symmetry.
There are also many examples of flows between interacting 
fixed points which are generated by various deformations.
These situations are discussed in later sections.

\subsection*{A. Sufficient conditions}

We first note that in part of the conformal window of
some models, the unitarity bound $r\geq 2/3$ fails for one or more composite 
operators of the chiral ring.
Then
the formulae (\ref{abcIR}) for infrared anomalies require correction
for the ensuing accidental symmetry.
Such cases are discussed separately in Sec. 4,
and we consider here models without accidental symmetry,
which necessarily have $r_i\geq 1/3$ for all fields of the microscopic 
theory.

The simplest way in which the positivity conditions can be
satisfied is if the contributions to $b_{IR}$, $c_{IR}$ and $a_{IR}$ in
(\ref{abcIR}), and to $a_{UV}-a_{IR}$ in
(\ref{aaflow}), are separately
positive for each contributing representation $R_i$. 
This leads
to the following sufficient conditions:

\begin{romanlist}
\item $b_{IR}>0$ if $r_i \leq 1$ for all chiral superfields
$\Phi^i$
\vspace*{2ex}

\item $c_{IR} >0$ if $1- \sqrt{5}/3 = .254 \leq r_i \leq 1$ or $r_i \geq
1 + \sqrt{5}/3=1.745$ for all $\Phi^i$
\vspace*{2ex}

\item $a_{IR}>0$ if $1- 1/\sqrt{3} = .423\leq r_i \leq 1$ or $r_i \geq
1 + 1/\sqrt{3} = 1.577$ for all $\Phi^i$
\vspace*{2ex}

\item $a_{UV} - a_{IR} \geq 0$ if $r_i \leq 5/3$ for all $\Phi^i$.
\end{romanlist}

In all of the models examined we find that in the part of the
conformal window where no accidental symmetry is required,

\begin{alphlist}
\item remarkably,
$r_i \leq 5/3$ for all renormalizable models, so the
$a$-theorem is always satisfied.

\item $1- \sqrt{5}/3 < r_i <1$ in all electric models without accidental
symmetry. 
Since electric and magnetic anomalies match in all
models, we have $b_{IR} >0$ and $c_{IR} >0$ on both sides of the duality.

\item $1- 1/\sqrt{3} < r_i <1$ is satisfied in part of the  
conformal window
of all theories, but not always. But the sufficient condition
is rather weak, and the positive contribution of the gauge
multiplet $a_{IR}$ always ensures $a_{IR}>0$ in the
non-accidental region.
\end{alphlist}

Thus, most 
of the positivity conditions,
especially the $a$-theorem, can be verified essentially by
inspection of the tables of $R$-charges presented in the
literature on the various models. 
Actually, in
many cases one
can prove that $r_i<5/3$ as a consequence of asymptotic freedom
in absence of 
accidental symmetry (i.e. when all $r_i\geq 1/3$).
Explicit check is then unnecessary.
We illustrate this in three simple situations 

i) For models with $N_f$ copies of a single irreducible
real representation $R$
(or $N_f$ copies of $R\oplus \overline{R}$),
one can see from the $S_{\mu}$ current in (\ref{Scurr})
that $\gamma^*=1-\frac{3T(G)}{N_fT(R)}$ (or 
$\gamma^*=1-\frac{3T(G)}{2N_fT(R)}$)
and asymptotic freedom gives $\gamma^*<0$ in both cases.
Thus $r=(2+\gamma^*)/3<2/3$.

ii) For renormalizable Kutasov-Schwimmer type models
the current (\ref{SScur}) immediately gives the same information,
$r<2/3$ for the fields in $R$ and $\overline{R}$
and $r_X=2/3$.

iii) We also consider models which have the same structure as 
magnetic SU$(N_c)$ 
SUSY QCD, namely $N_f$ fields $q$ in a real representation
$R'$ of a dual gauge group $G'$ (or $N_f$ fields $q$, $\tilde{q}$
in $R'\oplus \overline{R}'$) plus a gauge singlet meson field in the 
${\bf N_f}\otimes
{\bf N_f}$ (or ${\bf (1,N_f)}\otimes
{\bf (N_f,1)}$) representation of the flavor group
SU$(N_f)$ (or SU$(N_f)\times SU(N_f)$).
There is a superpotential $W=qMq$ (or $W=qM\tilde{q}$).
Here again one can inspect the gauge beta function 
(or the appropriate
$S_{\mu}$ current (\ref{SScur}))
and find $\gamma^*_q<0$ and $1/3\leq r_q<2/3$.
The superpotential then tells us that $r_M=2-2r_q$
satisfies $2/3<r_M\leq 4/3$ with the upper limit 
from unitarity without accidental symmetry.
Thus again $r_i<5/3$ for all fields. 

\subsection*{B. Flows between superconformal fixed points}

A conformal fixed point is characterized
by the values of $b$,
$c$ and $a$.
These values do not depend on the particular
flow which leads to or from this
conformal theory.
Therefore one may be interested in
a computation of the
flow $a_{UV} - a_{IR}$ for a theory which
interpolates between two
interacting conformal fixed points.
Such an interpolation may be obtained by deforming
a superconformal theory with a relevant operator
which generates an RG flow driving the theory to another
superconformal fixed point.
Since we know the conformal theories at both
ultraviolet and infrared limits of this
interpolating theory,
the computation simply requires subtraction of 
the end-point central charges.
In this case
we do not need to construct any $S$-current interpolating
between the ultraviolet and infrared conformal fixed points.
However it is interesting that in some cases one can
construct such an $S$-current and check directly
the value of the flow $a_{UV} - a_{IR}$.
We discuss below aspects of various types of deformations.
\vskip .1in

$\bullet$ {\it Mass deformations.}

The simplest case is a mass deformation.
Consider a conformal theory ($H$) characterized by
$a^H$,
$b^H$ and $c^H$
which contains a chiral superfield $\Phi$ in a real representation
of the gauge group (or a pair of
chiral superfields $\Phi$ and $\tilde{\Phi}$ in
conjugate representations).
Such a theory may be deformed by adding a gauge invariant
mass term
$W_m = \frac{m}{2} \Phi^2$ (or $W_m = m \Phi\tilde{\Phi}$).
We assume  that the heavy superfield $\Phi$
(or $\Phi$ and $\tilde{\Phi}$) decouples from the low-energy spectrum,
and that the resulting theory flows to a new conformal fixed point
with a smaller global symmetry group,
and characterized by the values $a^L$,
$b^L$ and $c^L$.
Since the heavy fields of the original theory do
not contribute to infrared anomalies, we have
$a_{IR} = a^L$,
$b_{IR} = b^L$, $c_{IR} = c^L$.
On the other hand the heavy fields
contribute to ultraviolet anomalies so that
$a_{UV}=a^H$,
$b_{UV}=b^H$ and $c_{UV}=c^H$.
Thus we have $a_{UV}-a_{IR}=a^H-a^L$.
As a result we expect that $a_{UV} >a_{IR}$.
This is indeed the case for all the models
that we have analysed.

One can obtain a simple analytic formula in the case
of an electric type theory with $N_f$ copies
of $R\oplus \overline{R}$ representation and 
no superpotential.
In this theory $r=1-T(G)/2N_f T(R)$ for the $N_f$ quarks
of the theory $H$.
We consider a mass deformation of $H$ which 
leaves
$N_f-n$ massless quarks in the theory $L$.  
These quarks have $r=1-T(G)/2(N_f-n)T(R)$.
Substituting these charges in the formula 
(\ref{aaflow}) we subtract with the result
$$a_H-a_L=\frac{9{\rm dim} R\, T(G)^3}{128\,T(R)^2}\left(-\frac{1}{N_f^2}+
\frac{1}{(N_f-n)^2}\right)>0.$$

In the special case of interpolation between
an ultraviolet free theory and an infrared
non-trivial conformal fixed point one can apply a more formal
argument.
In this case we consider the electric theory above with added mass term
for the $n$ massive quarks.
The unique $S_{\mu}$ current of this new theory is 
$$S_{\mu}=R_{\mu}+ \frac{1}{3}\left(
1-\frac{3T(G)}{2(N_f-n)T(R)}-\gamma_L\right)
K_{\mu}^L +\frac{1}{3}(1-\gamma_H)K_{\mu}^H,$$
where the superscripts $L$ and $H$ indicate Konishi currents
for the light and heavy quarks, respectively.
Thus $\gamma_H^*=1$ and $r_H=1$ so that the heavy quarks do not contribute
to $a_{IR}=a_L$ in (\ref{aaflow}).
For the light quarks $\gamma^*_L=
1-3T(G)/2(N_f-n)T(R)$ and $r_L=1-T(G)/2(N_f-n)T(R)$ which is exactly 
the correct value in the low-energy theory of $N_f-n$ flavors.
Thus the $S_{\mu}$ current analysis leads to the same value of 
$a_{IR}=a_L$ used above.

\vskip .1in
 
$\bullet$ {\it Higgs deformations.}

There are two qualitatively different types of Higgs
deformations.
The first is a deformation along flat directions of the potential
for the scalar fields.
Under such a deformation one generically breaks both the gauge and
flavor symmetries.
While the Goldstone bosons corresponding to the gauge symmetry
breaking are eaten by the Higgs mechanism, the Goldstone bosons  
of
the flavor symmetry breaking remain in the massless spectrum
of the theory.
Therefore these Goldstone bosons 
(and their superpartners) have to
be taken into account in the computation of the infrared values
of $a$, $b$ and $c$ of
the resulting theory.
It is implicitly assumed in the literature that
these Goldstone
superfields decouple from other light fields of the 
low-energy theory and
are free in the infrared.  
We thus assign $r=2/3$ to these fields.

In general the positivity of the flow $a_{UV}-a_{IR}$
under the Higgs deformations is
nontrivial evidence for the $a$-theorem.
In a simple situation of flow from the
higgsed ultraviolet free
theory to an infrared conformal fixed point
the positivity of $a_{UV}-a_{IR}$ 
follows from the following argument.
Let us consider
an asymptotically free theory $T$.
Let us also consider
an asymptotically free theory $T^{(1)}$ which
is a higgsed version of $T$ along a flat direction and
flows to a nontrivial
conformal theory in the infrared,
CFT$^{(1)}_{IR}$.
We are going to argue that the flow $a_{UV}(T^{(1)})-
a_{IR}^{(1)}> 0$.
We assume that there are $n$ Goldstone chiral superfields
that decouple
from the rest of the theory.
It is convenient
to define another asymptotically free theory $T^{(2)}$
which is just the theory $T^{(1)}$
with all massive fields dropped out plus $n$ free chiral
superfields.
Let us assume that the interacting part of
the theory $T^{(2)}$ is
also in its conformal window and
flows to a nontrivial conformal theory CFT$^{(2)}_{IR}$,
and the $a$-theorem is satisfied for this flow.
We have ${\rm CFT}^{(1)}_{IR}={\rm CFT}^{(2)}_{IR}
\oplus (n ~~{\rm free~~chiral~~superfields})$.
Therefore instead of the flow $T^{(1)}\to
{\rm CFT}^{(1)}_{IR}$ one can consider the two step flow
$T^{(1)}_{UV}\to T^{(2)}_{UV}\oplus 
(n~~{\rm free~~chiral~~superfields})
\to {\rm CFT}^{(1)}_{IR}$ (see Fig. 1).

\vskip .1in

\vbox{
\let\picnaturalsize=N
\def\picsize{3in}
\def\picfilename{diagramm.eps}
\ifx\nopictures Y\else{\ifx\epsfloaded Y\else\fi
\global\let\epsfloaded=Y
\centerline{\ifx\picnaturalsize N\epsfxsize \picsize\fi
\epsfbox{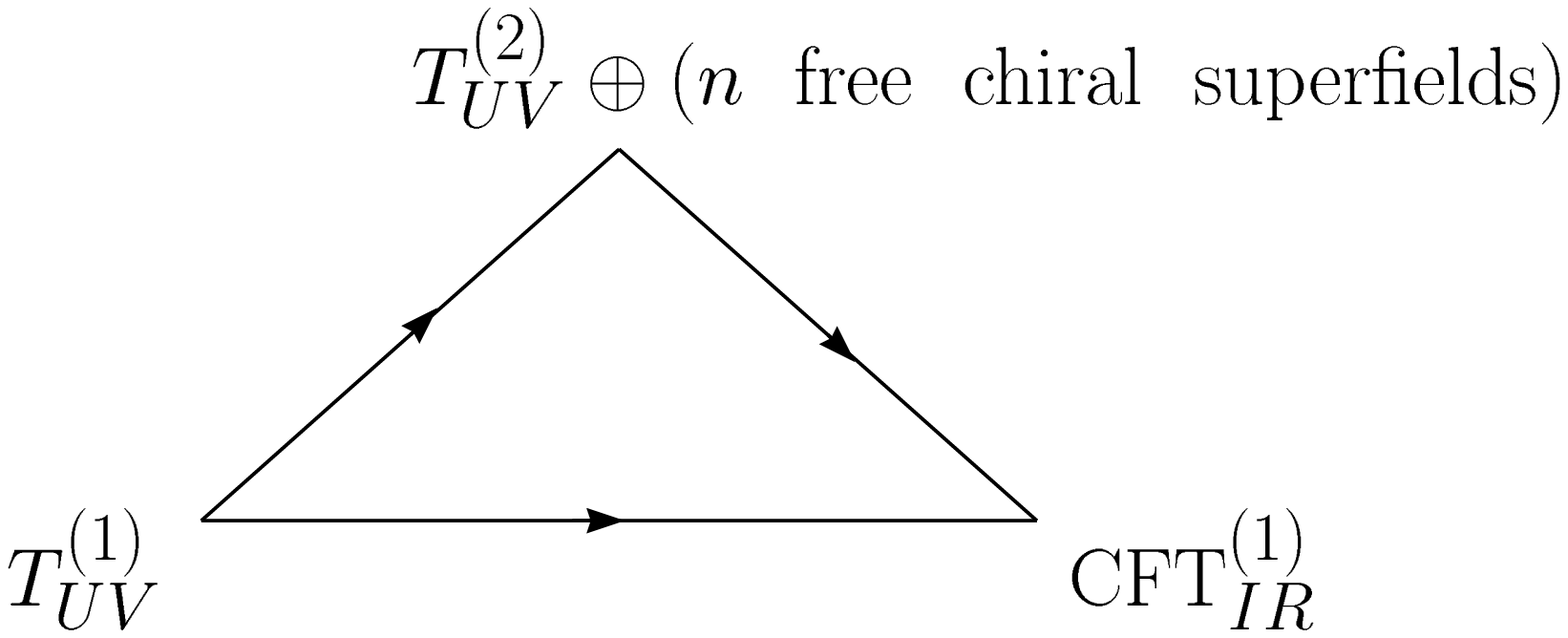}}}\fi

\centerline{Fig. 1. The diagram of flows under Higgs deformations.}
}
\vskip .1in
Since the $a$-theorem is trivially satisfied for the flow
$T^{(1)}_{UV}\to T^{(2)}_{UV}
\oplus (n~~{\rm free~~chiral~~superfields})$ we arrive at
the conclusion that $a_{UV}(T^{(1)})-
a_{IR}^{(1)}> 0$.

The second type of Higgs deformation
is the magnetic counterpart of a mass
term in the electric theory.
To be concrete we consider SU$(N_c)$
SUSY $QCD$ with electric quarks $Q_{\alpha}^i$ and anti-quarks
$\widetilde{Q}^{\alpha}_{i}$, where $ \alpha = 1 \ldots N_c$, and
$i = 1, \ldots, N_f$ are color and flower indices, respectively.
The magnetic theory has $G = SU (N_f - N_c)$ with quarks,
anti-quarks and meson $q^{\alpha}_{i}$, $\tilde{q}^i_{\alpha}$,
and $M^i_j$.
The mass perturbation $W_m = m Q^{N_f}_{\alpha}
\widetilde{Q}^{\alpha}_{N_f}$ in the electric theory is mapped to
$W_m = m M^{N_f}_{N_f}$ on the magnetic side \cite{emduality} so that
flavor symmetry is broken explicitly to SU$(N_f - 1)$.
Analysis \cite{emduality}
of the magnetic equations of motion shows that there is
a Higgs effect with $\langle q_{N_f} \tilde{q}^{N_f} \rangle \neq
0$, so the gauge group is broken to SU$(N_{f} - N_c -1)$.
The
spectrum contains massive fields plus the light fields of the
magnetic effective low energy theory with $G = SU(N_f - N_c -1)$
and $N_f -1$ flavors.
If this theory is still in its conformal
window, {\em i.e.} $N_{f} -1 > \frac{3}{2} N_c$, then $a_{IR}$ can be computed from (\ref{abcIR})
with the matter content and the gauge group of the low-energy theory.

As an example
one may consider a special case of the flow from the
higgsed ultraviolet free
theory to an infrared conformal fixed point.
It should be no surprise that there is also a formal argument
(based on a consideration of a conserved $S_{\mu}$ current)
since the conserved $R$-current on the electric side corresponds to
a conserved current on the magnetic.
One can verify that the
magnetic theory, with $W_m= mM^{N_f}_{N_f}$ has a unique set of
anomaly free $R$-charges.
There is an elaborate cancellation of
the contributions of heavy fields to the U$(1)_R$ and U$(1)_R^3$ anomalies,
and only the expected contributions from fields of the low energy
effective theory remain.

$\bullet$ {\it Deformations of superpotential.}

One can also consider more general deformations of the
superconformal theories by relevant operators.
A particular type of deformation is obtained by adding a relevant chiral
gauge invariant operator to the superpotential of
a superconformal theory.
As a result the deformed theory may flow to another fixed point
along the RG flow generated by the deformation.
In all renormalizable models
that we studied the induced flow of  $a$ is positive
but this is not true in non-renormalizable models (see Section 6).
Examples of interpolating flows are those between the $k$ and $k-1$
Kutasov-Schwimmer models which are discussed in Section 5.

\section{Accidental symmetries}

In this section we explain the computation of the infrared values of
$a$, $b$ and $c$ in the presence of accidental symmetry.
The appearance of accidental symmetry is associated with
an apparent violation of the unitarity bound $r\geq 2/3$
for a primary gauge invariant chiral composite field $M$.
The simplest hypothesis explored in the literature (for a review and 
discussion see ref. \cite{accid})
is that this signals that the field $M$ is
actually decoupled from the interacting part of the theory,
and becomes a free chiral superfield in the infrared \cite{accid}.

On the other hand the $R$ charge 
is equal to $2/3$ 
for a free chiral superfield, which contradicts
the result of computation with the $S_{\mu}$ current. 
A plausible explanation is that there is an additional anomaly free
global U$(1)$ generated by the spin-1 component $J^{(M)}_{\mu}$
of the composite superfield $\overline{M}M$.
The field $M$ is charged with respect to the current
$J^{(M)}_{\mu}$ but the other fields are not.
In this case the perturbative anomaly free
$S_{\mu}$ current can
mix with
the $J^{(M)}_{\mu}$ current under the RG flow
because the scaling dimension of the latter
tends to the canonical dimension 3 of a conserved current.
Thus the infrared $R$ current can be determined as an
infrared limit of a linear combination
\begin{eqnarray}
R_{\mu}^{IR}=S_{\mu}+ A_{\mu},
\label{acc}
\end{eqnarray}
where $A_{\mu}=\lambda J^{(M)}_{\mu}$.
The coefficient $\lambda$ is fixed by the condition that 
$R=2/3$ for the field $M$.

Assuming that this picture is correct one can easily compute
the infrared values of the central functions $a$, $b$ and $c$.
In the notation of Sec.~2, one has to compute the three point correlators
$\langle RRR\rangle_{IR}$ and $\langle RTT\rangle_{IR}$.
Substituting the expression (\ref{acc}) for $R_{\mu}$ into these
correlators
one has (the subscript $IR$ is omitted here)
\begin{eqnarray}
\langle RRR\rangle = \langle SSS\rangle +3\langle SSA\rangle
+3\langle SAA\rangle+
\langle AAA\rangle,~~~
\langle RTT\rangle = \langle STT\rangle +\langle ATT\rangle.
\end{eqnarray}
At this point we note that the correlators 
$\langle SSA\rangle$, $\langle SAA\rangle$, 
$\langle AAA\rangle$ and $\langle ATT\rangle$ are 
saturated by the free chiral field $M$
and hence they can be easily computed, {\em i.e.} we have
$$\langle SSA\rangle=\langle SSA\rangle_{\rm free}, ~~
\langle SAA\rangle=\langle SAA\rangle_{\rm free}, ~~
\langle AAA\rangle=\langle AAA\rangle_{\rm free},~~
\langle ATT\rangle=\langle ATT\rangle_{\rm free}.$$
Thus the correlators 
$\langle RRR\rangle_{IR}$ and $\langle RTT\rangle_{IR}$
can be rewritten as follows:
\begin{eqnarray}
\langle RRR\rangle_{IR}&=&
\langle SSS\rangle +\langle RRR\rangle_{\rm free}-
\langle SSS\rangle_{\rm free},\cr
\langle RTT\rangle_{IR}&=&
\langle STT\rangle+\langle RTT\rangle_{\rm free}-
\langle STT\rangle_{\rm free}.
\end{eqnarray}
As we explained in section 2 the central charges 
$a_{IR}$ and $c_{IR}$ are just given by
linear combinations of the correlators
$\langle RRR\rangle_{IR}$ and $\langle RTT\rangle_{IR}$.
We consider the case where there is one accidental U$(1)$ symmetry
for the gauge invariant composite superfield $M$ in 
an irreducible representation of the flavor group of dimension dim$M$
(more general cases can easily be handled).
The corrected infrared values of the central charges are
\begin{eqnarray}
a_{IR}&=&a_{IR}^{(0)}+
\frac{\rm dim~ M}{96}(2-3r_M)^2
(5-3r_M),\cr
 c_{IR}&=&c_{IR}^{(0)}+
\frac{\rm dim~ M}{384}
(2-3r_M)[(7-6r_M)^2-17].
\label{accid}
\end{eqnarray}
Here we denoted by $a_{IR}^{(0)}$ and $c_{IR}^{(0)}$ the 
expressions for $a$ and $c$ given by equations (\ref{abcIR}),
and $r_M$ stands for the $S$ charge of the chiral field $M$,
specifically the sum of the $S$ charges of its elementary constituents.
Since by assumption $r<2/3$ it is easy to see that the
correction to $a$ is always positive.
The correction to $c$ is positive at 
$r<(7 - \sqrt{17})/6\approx .479$ and
negative at  $.479\approx (7 - \sqrt{17})/6<r<2/3$.
In some models the accidental correction 
is required to make  $a_{IR}$ and $c_{IR}$ positive,
so the sign is important.

In general the formulas for the infrared values of  flavor central 
functions should also be corrected 
due to the presence of accidental symmetries.
The general formula for the corrected $b$ can be easily obtained
along the above lines and reads
$$
b_{IR}=b_{IR}^{(0)}+3
\,T^i_j T_i^j~\left(r_M-\frac{2}{3}\right) .
$$
Here we denoted by $b_{IR}^{(0)}$ the expression  
for $b_{IR}$ given in (\ref{abcIR}),
$T^i_j$ stands for the flavor generator associated with $b$.
The correction ${\rm dim}~ M~(r_M-2/3)$ is always negative.

{\it Deformations of conformal fixed points with accidental 
symmetry.}
In the following we test various examples of superconformal 
models and flows between them.
In particular we will consider flows from 
superconformal 
models with accidental symmetries taken as 
an ultraviolet fixed point to different infrared fixed points.
Such a flow may be generated by appropriate deformation 
of the ultraviolet theory  
with a relevant operator.
It is important that the ultraviolet theory has to be taken 
together with the free chiral fields generating the accidental symmetry.
In fact the deformation of the ultraviolet theory by a relevant
operator generates a non-trivial
coupling of the interacting part of the UV theory to
the accidental chiral superfields.
This turns out to be important for positivity of $a_{UV}-a_{IR}$.

\section{Examples of models with uniquely defined
$S$ current and the flows}

In this section we give detailed results for
the models that we have analyzed.
We mainly focus on subtleties met in 
the computations of the infrared values of $a$ and $c$.

\subsection{Models with one type of irreducible representation}

This class of models includes the SU$(N_c)$ series,
SO$(N_c)$ series \cite{emduality},
Sp$(2N_c)$ series \cite{Intsei},
Pouliot Spin(7) model \cite{pouliot},
Distler-Karch models with exceptional groups \cite{distler}.
\bigskip

$\bullet$ Seiberg's QCD with $G=SU(N_c),~SO(N_c)$ with $N_f$, 
and Sp$(2N_c)$
with $2N_f$ fundamentals.
Conformal windows are
$3N_c/2< N_f(SU) <3N_c$,
$3(N_c-2)/2<N_f(SO)<3(N_c-2)$, $3(N_c+1)/2<N_f(Sp)<3(N_c+1)$.
There are no accidental symmetries.
Since all $R$ charges obey $r\leq 5/3$ 
we always have
$\Delta a=a_{UV}-a_{IR}>0$
for the flows from the free ultraviolet to 
conformal fixed points.
The results of our computations are given Table 1.
It should be noted that all flows vanish quadratically in the respective
weakly coupled limits of electric and magnetic theories.
This agrees with the discussion of the perturbative limit at the end 
of Sec. 2.

\vskip .2in
\vbox{
Table 1. Flows from  UV free theories to Seiberg's conformal QCD.
\hskip .1in

\begin{tabular}{|c|c|c|}
\hline
 Gauge group & $a_{UV}-a_{IR}$ in electric theory & $a_{UV}-a_{IR}$ in magnetic theory  \\
\hline
 SU$(N_c)$ & ${N_fN_c\over 48}\left(1-\frac{3N_c}{N_f}\right)^2
\left(2+\frac{3N_c}{N_f}\right)$ & 
$\frac{1}{12}\left(1-\frac{3N_c}{2N_f}\right)^2(3N_c^2+4N_cN_f+3N_f^2)$  \\
\hline
 SO$(N_c)$ & $\frac{N_c( -6+2N_f+3N_c)
( 6 + N_f - 3N_c)^2}{96N_f^2}$ & 
$\frac{N_c( -6-2N_f+3N_c)^2
(3N_f^2-6N_c+4N_cN_f+3N_c^2 )}{96N_f^2}$  \\
\hline
 Sp$(2N_c)$ & $\frac{( -3 + N_f - 3N_c)^2
N_c( 3 + 2N_f + 3N_c)}{24N_f^2}$ & 
$\frac{(3 -2 N_f+3N_c)^2
(3N_f^2 + 3N_c + 4N_cN_f+3N_c^2)}{24N_f^2}$  \\
\hline
\end{tabular}
}
\hskip .2in

The models considered below have non-renormalizable 
magnetic versions.
Therefore we discuss only the electric versions that are
renormalizable.
The results of our computations are given in Table 2.
Aspects of the RG flows of non-renormalizable
theories are considered in the next section.

$\bullet$ Spin(7) Pouliot model with $N_f$ spinors {\bf 8},
$Q_i$.
Conformal window: $7\leq N_f\leq 14.$
We have in the infrared
$r_{\bf 8}^{IR}=1-5/N_f$.
There is an accidental symmetry at $N_f=7$ due to decoupled $QQ$ singlet.
In Table 2 we separated the accidental corrections to $a_{IR}$ and $c_{IR}$ from the regular ones. 
Note that the correction to $c_{IR}$ turns out to be negative.

$\bullet$ $G_2$ with $N_f$ {\bf 7}.
Conformal window: $6\leq N_f\leq 11.$
We have $R^{IR}_{\bf 7}=1-4/N_f$.
The accidental symmetry point appears at $N_f=6$
where $QQ$ has $r=2/3$ and hence it is free.
Therefore there are no accidental corrections to the central charges.

$\bullet$ $E_7$ Distler-Karch model: 4 fundamentals {\bf 56}, $Q_i$;
$r_Q=1/4$.

$\bullet$ $E_6$ Distler-Karch model (I): 6 fundamentals {\bf 27},
$Q_i$; $r_Q=1/3$.

$\bullet$ $E_6$ Distler-Karch model (II): 
$3\times ({\bf 27}+{\bf \overline{27}})$ fundamentals $Q_i$;
$r_Q=1/3$.

$\bullet$ $F_4$ Distler-Karch model: 5 fundamentals {\bf 26}, $Q_i$;
$r_Q=2/5$.

$\bullet$ $F_4$ Distler-Karch model: 4 fundamentals {\bf 26}, $Q_i$;
$r_Q=1/4$.
There is an accidental symmetry associated with decoupling
of meson fields $M_{ij}=Q_iQ_j$.
In Table 2 we separated the accidental corrections to $a_{IR}$ and $c_{IR}$ from the regular ones. 
Again the correction to $c_{IR}$ turns out to be negative.

$\bullet$ Spin(8) Distler-Karch model: 
$4\times ({\bf 8}_v+{\bf 8}_c+{\bf 8}_s)$ fundamentals $Q$;
$r_Q=1/2$.

\hskip 1in

\vbox{
Table 2. The infrared $a$ and $c$ charges, and flows from the
ultraviolet free theory to conformal fixed points. 
\hskip .1in

\hskip -.2in\begin{tabular}{|c|c|c|c|}
\hline
 Model & $a_{IR}$ & $c_{IR}$ &  \vbox{\hbox{electric theory,} \hbox{$a^{\rm free}_{UV}-a_{IR}$}} \\
\hline
 \vbox{\hbox{Spin(7) with $N_f>7$ spinors {\bf 8}}\hbox{no accidental symmetry}} & 
${123\over16}-{1125\over 4N_f^2}>0$ & ${71\over 8}-
{1125\over 4N_f^2}>0$ & 
$\frac{N_f}{12}\left(1-\frac{15}{N_f}\right)^2\left(2+\frac{15}{N_f}\right)$  \\
\hline
 \vbox{\hbox{Spin(7) with $N_f=7$ spinors {\bf 8}}\hbox{accidental 
symmetry}}
 & ${1527\over 784}+\frac{23}{168}=\frac{4903}{2352}$ & ${1229\over 392}-\frac{13}{84}=\frac{3505}{1176}$ & ${3551\over 1176}$  \\
\hline
 \vbox{\hbox{$G_2$ with $7\leq N_f\leq 11$ in {\bf 7}}
\hbox{no accidental symmetry}} & 
${21\over 4}-{126\over N_f^2}>0$ & ${49\over 8}-{126\over N_f^2}>0$ & 
$\frac{7N_f}{48}\left(1-\frac{12}{N_f}\right)^2\left(1+\frac{6}{N_f}\right)$  \\
\hline
 $E_7$ with 4 fundamentals {\bf 56} & ${903\over 64}$ & ${1043\over 64}$ & ${2975\over 192}$  \\
\hline
 $E_6$ with 6 fundamentals in {\bf 27} & $\frac{45}{4}$ & $\frac{105}{8}$ & $\frac{27}{4}$  \\
\hline
 $E_6$ with matter in
$3\times ({\bf 27}+{\bf \overline{27}})$ & ${45\over 4}$ & ${105\over 8}$ & ${27\over 4}$  \\
\hline
 $F_4$ with $N_f=5$ in {\bf 26} & ${1833\over 200}$ & ${1079\over 100}$ & ${247\over 75}$  \\
\hline
 \vbox{\hbox{$F_4$ with $N_f=4$ in {\bf 26}}
\hbox{accidental symmetry}} & ${{1209}\over {256}}+{7\over {48}}=
{{3739}\over {768}}$ & ${{1625}\over {256}}
-{1\over {48}}={{4859}\over {768}}$ & ${{5413}\over {768}}$  \\
\hline
 \vbox{\hbox{Spin(8) with matter in}\hbox{
$4\times ({\bf 8}_v+{\bf 8}_c+{\bf 8}_s)$}}
 & ${51\over 8}$ & ${61\over 8}$ & ${7\over 8}$  \\
\hline
\end{tabular}
}
\hskip 1in

\subsection{Deformations}

$\bullet$ Deformations of SU$(N_c)$, SO$(N_c)$ and
Sp$(2N_c)$ Seiberg QCD models.
Higgsing of the Seiberg superconformal models corresponds to 
$N_c,~N_f\to N_c'=N_c-1, ~N_f'=N_f-1$.
The infrared theory has $2(N_f-1)$ decoupled 
Goldstone gauge singlets for
SU$(N_c)$ and Sp$(2N_c)$ models and $N_f-1$ for SO$(N_c)$. 

1. Consider first the SU$(N_c)$ theory.
In the region $3N_c/2<N_f\leq 3N_c-3$ both the 
ultraviolet and infrared theories are in their conformal
windows
and we have 
$$\Delta~a={{1 - N_f}\over {24}}+ {3(2N_c-1)\over 8} - 
  {{9\,{N_c^4}}\over {16\,{N_f^2}}}+
  {{9\,{(N_c-1)^4}}\over {16\,{(N_f-1)^2}}}>0.$$
In the cases $N_f=3N_c-1$,  $3N_c-2$ the infrared theory is free
since $N_f'=3N_c'+1$ and $N_f'=3N_c'$ respectively.
The infrared value $a_{IR}$ is then computed using 
$r=2/3$ for all chiral superfields of the low-energy 
$N_c'$, $N_f'$ theory and the Goldstone fields.
The results are
$$\Delta~a={{-9 + 76N_c - 210\,{N_c^2} + 180\,{N_c^3}}\over 
   {48\,{{\left( -1 + 3N_c \right) }^2}}}>0,~~~{\rm and}~~~
\Delta~a=
{{\left( -2 + 5N_c \right) \,
     \left( 6 - 19N_c + 12\,{N_c^2} \right) }\over 
   {16\,{{\left( -2 + 3N_c \right) }^2}}}>0.$$

2. Consider the SO$(N_c)$ theory.
In the region $3(N_c-2)/2<N_f\leq 3N_c-8$ both the 
ultraviolet and infrared theories are at their conformal fixed points
and we have 
$$
\Delta~a={{1 - N_f}\over {48}}+\frac{3}{32}
\left[
2N_f+8(N_f-N_c+2)-9\left(\frac{N_c}{N_f}-\frac{N_c-1}{N_f-1}
\right)(N_f-N_c+2)^2+\right.$$
$$\left.
3\left(\frac{N_c}{N_f^2}-\frac{N_c-1}{(N_f-1)^2}
\right)(N_f-N_c+2)^3
\right]>0
$$
In the cases of $N_f=3N_c-7$, $3N_c-8$
(in the latter case we limit ourselves to $N_c\geq 4$
for the ultraviolet theory to be in the
conformal window) the infrared theory is free
so that respectively
$$\Delta~a={{-882 + 1756N_c - 1011\,{N_c^2} + 180\,{N_c^3}}\over 
   {96\,{{\left( -7 + 3N_c \right) }^2}}}>0,~~~~
\Delta~a={{-192 + 372N_c - 193\,{N_c^2} + 30\,{N_c^3}}\over 
   {16\,{{\left( -8 + 3N_c \right) }^2}}}>0.$$

3. Consider the Sp$(2N_c)$ theory.
In the region $3(N_c+1)/2<N_f\leq 3N_c+1$ both the 
ultraviolet and infrared theories are at their conformal fixed points
and we have 
$$\Delta~a={{1 - N_f}\over {24}}+\frac{1}{32}
\left[
6(3-4N_f)+96(N_f-N_c-1)-108\left(\frac{N_c+1}{N_f}-\frac{N_c}{N_f-1}
\right)(N_f-N_c-1)^2+\right.$$
$$\left.
36\left(\frac{N_c+1}{N_f^2}-\frac{N_c}{(N_f-1)^2}
\right)(N_f-N_c-1)^3
\right]>0.
$$
In the cases of $N_f=3N_c+1$, $3N_c+2$ the infrared theory is free
so that respectively
$$\Delta~a={{-3 - 16N_c + 41\,{N_c^2} + 138\,{N_c^3}}\over 
   {16\,{{\left( 1 + 3N_c \right) }^2}}}>0,~~~{\rm and}~~~
\Delta~a=
{{-28 + 86N_c + 471\,{N_c^2} + 414\,{N_c^3}}\over 
   {48\,{{\left( 2 + 3N_c\right) }^2}}}>0.$$
The mass deformations obviously respect the $a$-theorem because
$\partial a/\partial N_f>0$ in all cases
(see explicit computation in Sec. 3).

$\bullet$ Deformations of Spin(7) Pouliot model.

First consider the higgsing of the Spin(7) Pouliot model
with $7\leq N_f\leq 14$ fundamentals to the
$G_2$ model with $N_f-1$ fundamentals and $N_f-1$ Goldstone superfields.

In the region $8\leq N_f\leq 14$ there are no accidental symmetries
either in the ultraviolet or in the infrared.
Thus we have
$r_{\bf 8}^{UV}=1-5/N_f$, $r^{IR}_{\bf 7}=1-4/(N_f-1)$ and 
$r^{IR}_{\bf 1}=2/3.$
The flow is
$$
\Delta a={1\over 144N_f^2(N_f-1)^2}
(13500-27000N_f+7523N_f^2-141N_f^3+69N_f^4+N_f^5)>0.$$

Note that for $N_f=13,14$ the infrared $G_2$ theory is free.
In this case we have
$$\Delta a~(N_f=13)={{3781}\over {2704}},~~~
\Delta a~(N_f=14)={{859}\over {588}}.$$
For $N_f=7$ the UV theory has an accidental symmetry.
One has
$$\Delta a=\frac{1945}{2352}.$$

Mass deformations.
By giving a mass to one of the flavors one can generate 
the flow $N_f\to N_f-1$.
Obviously, $a_{UV}-a_{IR}=a~(N_f)-a~(N_f-1)>0$.

\vbox{
$\bullet$ The results of computations for the flows induced by Higgsing 
of Distler-Karch superconformal models are given in Table 3.

\vskip .1in
\centerline{Table 3. Higgs deformations of Distler-Karch models.}
\centerline{
\begin{tabular}{|c|c|c|c|}
\hline
Higgsing & $F_4\to Spin (8)$ & $E_6\to F_4$ & $E_7\to E_6$  \\
\hline
$a_{UV}-a_{IR}$ & ${2623\over 300}$ & ${2377\over 1200}$ & ${175\over 64}$  \\
\hline
\end{tabular}}
}
\vskip .1in

$\bullet$ Mass deformation of $F_4$ model \cite{distler}.
By giving a mass to one of flavors the theory with $N_f=5$
is driven to 
a new conformal fixed point with $N_f=4$ flavors $Q_i$ and $r_Q=1/4$.
The theory has an accidental symmetry associated with decoupling 
of the 16 mesons $M_{ij}=Q_iQ_j$, $r_M=2/3$.
For the flow from $N_f=5$ to $N_f=4$ we have
$$\Delta a={{85693}\over {19200}}.$$

\subsection{Models with two types of irreps
with uniquely determined $S$ current}

This set of models includes
those given in refs. \cite{kut} for SU,
\cite{intr,leigh} for SO and Sp 
gauge groups.
We discuss in detail only the SU Kutasov-Schwimmer models
and the Pouliot Spin(7) model with $N_c+4$ flavors in $\bf 8$
and singlets \cite{pouliot}. 
For these models we discuss also various flows between conformal 
fixed points.

\bigskip

$\bullet$ Consider the Kutasov-Schwimmer model \cite{kut}
with the SU$(N_c)$ gauge group,
$N_f$ flavors of quarks, $Q$ and $\tilde{Q}$ in the fundamental,
and a chiral superfield $X$ in the adjoint representation.
The superpotential is $W=X^{k+1}$.
The $R$-charges are
given in Table 4.

\vskip .1in

\centerline{Table 4. Matter content of Kutasov-Schwimmer models.}

\vskip .1in

\centerline{
\begin{tabular}{|c||c||c|c|c|}
\hline
~& SU$(N_c)$ & SU$(N_f)_Q$ & SU$(N_f)_{\widetilde{Q}}$ & U$(1)_R$  \\ 
\hline
$Q$ & $\Yfund$ & $\Yfund$ &~ & $1-\frac{2N_c}{(k+1)N_f}$ \\
\hline
$\widetilde{Q}$ &$ \overline{\Yfund}$ & ~& 
$\Yfund $& $1-\frac{2N_c}{(k+1)N_f}$ \\
\hline
$X $& adj &~ & ~& $\frac{2}{k+1}$\\
\hline
\end{tabular}}

\vskip .1in

The theory has a dual
with gauge
group SU($kN_f-N_c$), with $N_f$ flavors of ($\Yfund+\overline{\Yfund}$), an
adjoint and gauge singlets.  The conformal window is presumed to be the
region in $N_f,\,N_c$ where both the electric and magnetic theories are
asymptotically free, 
$${2N_c\over 2k-1}<N_f<2N_c.
$$
There is an  accidental symmetry in the range
$$
{2N_c\over 2k-1}<N_f\leq {3N_c\over k+1},
$$
where it corresponds to $QX^j\tilde{Q}$
out of the unitary region for one or more  values of $j$.
This accidental symmetry may appear in the conformal window for any
$k\geq 2$ (and sufficiently large $N_c$).
In particular, for $k=2$ it appears for $N_f\leq N_c$,
and for $k=3$ it appears for $N_f\leq 3N_c/4 .$

The only explicitly renormalizable Kutasov-Schwimmer 
model corresponds to $k=2$,
and it is studied below. 
The $k=3$ theory can be made renormalizable in part of its conformal 
window, and this is discussed in section 6.

In the case of absence of the accidental symmetry 
we may use eqs. (\ref{abcIR}).
We have
\begin{eqnarray}
a_{IR}&=&{9\over 32}
\left[\left(\left({2\over k+1}-1\right)^3+1\right)(N_c^2-1)-
{16\over (k+1)^3}{N_c^4\over N_f^2}+ 
{2\over 3}{N_c^2+1\over k+1}\right],\nonumber\\
c_{IR}&=&{9\over 32}
\left[\left(\left({2\over k+1}-1\right)^3+1\right)(N_c^2-1)-
{16\over (k+1)^3}{N_c^4\over N_f^2}+ 
{10\over 9}{N_c^2+1\over k+1}\right],\nonumber\\
\Delta a &=&-{9\over 32}\left[\left(\left({2\over k+1}-1\right)^3+
{7\over 27}\right)(N_c^2-1)-
{16\over (k+1)^3}{N_c^4\over N_f^2}+ {2\over 3}{N_c^2+1\over k+1}-
{4N_f\,N_c \over 27}\right].
\label{acKS}
\end{eqnarray}
It is obvious that 
$\Delta a>0$ in the conformal window since for all 
chiral fields  
$r_{IR}\leq 5/3$.

At $k=2$ we have 
$$\Delta a={N_c\over 24} \left(1-\frac{2N_c}{N_f}\right)^2
\left(N_c+{N_f}\right) \geq 0.$$
Note also that the first two equations in (\ref{acKS}) 
agree with the results for 
Seiberg's QCD at $k=1$.

We now consider the contribution of the accidental symmetry.  
We concentrate 
on the renormalizable case, $k=2$.  
In the region $2N_c/3<N_f\leq N_c$,
the meson operator $M=Q\widetilde{Q}$
has $r_M=2(1-2N_c/3N_f)<2/3$,
so there is an accidental correction to $c_{IR}$ and $a_{IR}$
(\ref{acKS}).
First we note that for large $N_c$ and $N_f \approx 2N_c/3$,
the previous formulae (\ref{acKS}) for the $k=2$ central charges 
without accidental contributions give
$$c^{(0)}_{IR}=-\frac{1}{6},~~~ a^{(0)}_{IR}=-\frac{1}{24}N_c^2$$
and are negative.
This is not surprising
since the theory is effectively nonunitary if the decoupling
of the meson field is not taken into account.  
Positivity is restored by the accidental contribution, and
this is an interesting check on the entire hypothesis
of accidental symmetry.
The sum of (\ref{acKS}) and the accidental correction (\ref{accid})
are 
\begin{eqnarray}
a_{IR}&=&-{3\over {16}} - {{{N_f^2}}\over 6} + N_fN_c - 
  {{7 {N_c^2}}\over 6} + {{2 {N_c^3}}\over {3N_f}} - 
  {{{N_c^4}}\over {6{N_f^2}}}>0,\cr
c_{IR}&=&-{1\over 8} - {{{N_f^2}}\over {12}} + {{11N_fN_c}\over {12}} - 
  {{9\,{N_c^2}}\over 8} + {{2\,{N_c^3}}\over {3N_f}} - 
  {{{N_c^4}}\over {6\,{N_f^2}}}>0.
\end{eqnarray}
We note that intrinsically positive accidental corrections to
$a_{IR}$ decrease $a_{UV}-a_{IR}$ and thus tend
to threaten the $a$-theorem.
Nevertheless we find that with the accidental contribution included
\begin{equation} \Delta a= 
{{11N_c^2}\over 8}+{N_c^4\over {6\,{N_f^2}}}-{2N_c^3\over {3N_f}}- 
   {{23N_fN_c}\over {24}}+{{{N_f^2}}\over 6}\geq 0. \end{equation}
The contribution of
the accidental symmetry to $b$ is always negative.  
However, we find that
all positivity conditions, including $b>0$,
are satisfied for $N_f$, $N_c$ in the accidental
window.
For example, for the central charge of
the SU$(N_f)_Q$ current we find for $k=2$
\begin{eqnarray}
b_{IR}=
\frac{4}{3N_f}\left( 2{N_f^2} - 2N_fN_c + 
{N_c^2} \right) >0.
\end{eqnarray}

\vskip .2in
$\bullet$ Deformations of Kutasov-Schwimmer 
superconformal models.

i. {\it Consider now the $k\to k-1$ interpolation.}

The simplest case is to consider $W={\rm Tr} X^{k+1}+{\rm Tr} X^k$
with $\langle X\rangle =0$ and
unbroken gauge group \cite{kut}.

As mentioned above our approach is not expected to work
for $k>3$ where there is no 
renormalizable description of the theory.
For $k=3$ and $N_c\geq N_f$ there is a renormalizable
description
that will be discussed in the next section.
Here we just note that in this region in the absence of accidental symmetries
the central charges 
are given by eqs. (\ref{acKS}) at $k=3$.
In particular at $N_c= N_f\geq 3$ (the only point in the 
renormalizable conformal window with no accidental 
symmetry) and for the flow $k=3\to k=2$ we have 
$$
\Delta a={7\over {768}}+{{43\,{N_c^2}}\over {768}}>0.$$
At $N_c=2$ the $k=2$ Kutasov-Schwimmer model is not defined since 
${\rm Tr} X^3=0$.
Instead one can consider the flow from 
the $k=3$ $N_f=2$ fixed point in the ultraviolet 
to 
$k=1$, i.e to Seiberg's SU$(2)$ SUSY QCD
with $N_f=2$ flavors.
This infrared theory is confining and the flat directions are lifted
due to non-perturbative quantum corrections
\cite{moduliS}.
As a result the SU$(4)$ global symmetry
is broken to Sp$(4)$.
The infrared low-energy theory is described by 5 free
chiral superfields with $r=2/3$.
Thus we have 
$$\Delta a={{451}\over {768}}.
$$

{\it Accidental symmetry.}
Consider first the $k=3 \to k=2$ flow with
an accidental symmetry 
($Q\tilde{Q}$) in the IR and none in the UV.
This corresponds to $3N_c/4<N_f<N_c$.
We have
$$\Delta a=-{3\over {256}} + {{{N_f^2}}\over 6} - N_fN_c + 
  {{1121\,{N_c^2}}\over {768}} - {{2\,{N_c^3}}\over {3N_f}} + 
  {{37\,{N_c^4}}\over {384\,{N_f^2}}}>0.$$
In the region $2N_c/3 <N_f <3N_c/4$
there is an accidental symmetry ($Q\tilde{Q}$)
in both the IR and UV, and the above expression has to be corrected.
Obviously, $\Delta a>0$ since the accidental contribution to 
the UV theory is positive.

For $N_f\leq 2N_c/3$ the infrared theory is the free magnetic 
$k=2$ theory \cite{kut} (again we must consider $N_c\geq 3$).
The value of $a_{IR}$ can be computed by assigning $r=2/3$
to all chiral superfields of the magnetic theory.
In the region $6N_c/11\leq N_f<2N_c/3$ the ultraviolet theory has 
only one accidental symmetry ($Q\tilde{Q}$)
and we have
$$\Delta a =-{{51}\over {256}} + {{211\,{N_f^2}}\over {1152}} + 
  {{5\,{N_f^4}}\over {288}} - {{3N_fN_c}\over 4} - 
  {{{N_f^3}N_c}\over {64}} + {{291\,{N_c^2}}\over {256}} + 
  {{5\,{N_f^2}\,{N_c^2}}\over {1152}} - 
  {{9\,{N_c^3}}\over {32N_f}} - {{9\,{N_c^4}}\over {128\,{N_f^2}}}>0.$$
For $N_f <6N_c/11$ there is an additional accidental symmetry ($QX\tilde{Q}$)
so that $a_{UV}$ increases and again $\Delta a >0.$

Consider the $k=2\to k=1$ flow.
The infrared theory is just Seiberg's QCD in the conformal phase.
There is no accidental symmetry in the physical window 
in the IR, for $N_f\geq N_c$.
In the region $N_f\geq 3N_c/2$ the IR theory is
at the conformal fixed point
we have
$$\Delta a=-{1\over {48}}-{{{N_c^2}}\over {24}}+ 
  {{19\,{N_c^4}}\over {48\,{N_f^2}}}>0.$$
For $N_f\leq 3N_c/2$ the IR theory is free.
By using the magnetic description of Seiberg's QCD 
to compute $a_{IR}$ we get
$$\Delta a={7\over {48}}-{1\over {48\,{N_c^2}}}- 
{N_c^2\over {6\,{N_f^2}}}+ 
  {{5\,N_f}\over {12N_c}}-{{{N_f^2}}\over 4N_c^2}>0.$$

ii. {\it Higgsing by $\langle X\rangle\neq 0$.}

We now consider the non-trivial stationary point
of the deformed superpotential \cite{kut}
that corresponds to the
breaking SU$(N_c)\to SU(N_c-1)\times U(1)$.
Consider $N_c\to N_c-1$ and $k\to k-1$, $k-2$
and $k=2,3$.    

-- The flow $k=2\to k=1$, $N_c\to N_c-1$ ($N_c\geq 3$).
The infrared theory is Seiberg's QCD 
(plus $2N_f$ free chiral superfields)
so that we have to consider only
$N_c\leq N_f <2N_c$.
At $N_c\leq N_f\leq 3N_c/2$ the infrared theory is confining 
and can be described by the
free magnetic theory with $r=2/3$ for all chiral superfields.
In this case we have 
$$\Delta a=-{{19}\over {48}} - {{11N_f}\over {24}} - {{{N_f^2}}\over 4} + 
  {{3N_c}\over 8} + {{5N_fN_c}\over {12}} + 
  {{7\,{N_c^2}}\over {48}} - {{{N_c^4}}\over {6\,{N_f^2}}}>0.$$
At $3N_c/2<N_f<2N_c$ the infrared theory is in the non-Abelian 
Coulomb phase (plus $2N_f$ free chiral superfields) 
and we have 
$$\Delta a=-{7\over {12}} + {9\over {16\,{N_f^2}}} - {N_f\over {24}} + 
  {{3N_c}\over 4} - {{9N_c}\over {4\,{N_f^2}}} - 
  {{{N_c^2}}\over {24}} + {{27\,{N_c^2}}\over {8\,{N_f^2}}} - 
  {{9\,{N_c^3}}\over {4\,{N_f^2}}} + 
  {{19\,{N_c^4}}\over {48\,{N_f^2}}}>0.$$

-- The flow $k=3\to k=2$, $N_c\to N_c-1$
($N_c\geq 4$).  
The infrared
theory is in its non-Abelian Coulomb phase.
If $N_c=N_f$ then there are no accidental symmetries either in the
UV or IR. 
Thus we have
$$\Delta a={{125}\over {256}} + {1\over {6\,{N_c^2}}} - {2\over {3N_c}} - 
  {N_c\over {24}} + {{43\,{N_c^2}}\over {768}}>0.$$
In the region $3N_c/4\leq N_f< N_c$ there is an accidental symmetry
in the IR and none in the UV.
We have
$$
\Delta a=
{{253}\over {256}} + {1\over {6\,{N_f^2}}} + {2\over {3N_f}} + 
  {{23N_f}\over {24}} + {{{N_f^2}}\over 6} - {{7N_c}\over 3} - 
  {{2N_c}\over {3\,{N_f^2}}} - {{2N_c}\over N_f} - N_fN_c + 
  {{1121\,{N_c^2}}\over {768}} + {{{N_c^2}}\over {{N_f^2}}} + $$
$$
  {{2\,{N_c^2}}\over N_f} - {{2\,{N_c^3}}\over {3\,{N_f^2}}} - 
  {{2\,{N_c^3}}\over {3N_f}} + {{37\,{N_c^4}}\over {384\,{N_f^2}}}>0.
$$
In the region $2N_c/5<N_f<3N_c/4$ both the UV and IR theories 
have accidental symmetries so that both $a_{UV}$ and $a_{IR}$ increase. 
This accidental contribution in the UV is crucial for  
$\Delta a>0$ in this region.

-- The flow $k=3\to k=1$, $N_c=3$.
We have to consider $N_f=2,3$.
In both cases the infrared theory is Seiberg's SU$(2)$ QCD 
with $N_f$ flavors in 
the confining phase.
At $N_f=2$ the infrared theory contains just 5 free 
chiral superfields with $r=2/3$.
The UV theory has an accidental symmetry ($Q\tilde{Q}$).
Thus we have 
$$\Delta a={{1453}\over {1536}}.$$
At $N_f=3$ the infrared
theory is described by nine free mesons
and two baryons ($r=2/3$), and
$$\Delta a={{605}\over {384}}.$$

-- The flow $k=3\to k=1$, $N_c=2$.
We have to consider $N_f=2$.
The infrared theory is a U$(1)$ gauge theory 
with 2 flavors,
which is infrared free.
We have  
$$\Delta a=\frac{323}{768}.$$

iii. {\it Higgsing along flat directions.}

One can change $N_c\to N_c-1$ and $N_f\to N_f-1$
by turning on $\langle Q_{N_f}\rangle =\langle \tilde Q_{N_f}
\rangle \neq 0$.
One can show 
for sufficiently large $k$ which correspond to 
non-renormalizable models the $a$-theorem is violated 
due to the  negative contribution of Goldstone superfields.
However $\Delta a>0$ in the renormalizable cases $k\leq 3$.
This is the first observed problem with the $a$-theorem and
we discuss it in Sec. 6 after further study of non-renormalizable cases.

iv. {\it Massive deformations.}

By adding a mass term to one of the flavors one can reduce
$N_f\to N_f-1$.
This obviously gives $\Delta a>0$ since 
$\partial a/\partial N_f <0.$

\vskip .1in

$\bullet$ Spin(7) Pouliot model
with $N_c+4$ spinors {\bf 8}, $q_i$,
with $r_q=1-5/(N_c+4)$,
singlets $M_{\{i,j\}}$ with $r_M=10/(N_c+4)$.
There is a superpotential $Mq\tilde q$.
We have
$$c={-398 + 87\,N_c + 74\,N_c^2 + 
     38\,N_c^3 - N_c^4\over 
16\,( 4 + N_c)^2},~
a={3(\, -308 + 42\,N_c + 44\,N_c^2 + 
       23\,N_c^3 - N_c^4) \over 
32\,( 4 + N_c)^2}.$$
\vskip .1in

For the flow from the ultraviolet free theory
to the conformal fixed point we have
$$
\Delta a=
{{ {( -11 + N_c)^2}\,
       \left( 42 + 23\,N_c + 5\,{{N_c}^2} \right)  
 }\over {48\,{( 4 + N_c)^2}}}>0.
$$
{\it Higgs deformation of the model}:
one can check that 
$\Delta a>0$ under the flow $N_c\to N_c-1$ 
in the conformal
window ($N_c\leq 10$).


\section{Nonrenormalizable Kutasov-Schwimmer Models}

In this section
we shall study flows 
of central charges in models which are
non-renormalizable as fundamental theories
with Kutasov-Schwimmer models for
$k\geq 3$ as examples.
It is open to question whether our method is correct for 
non-renormalizable theories, but we analyze
the data first and then discuss the situation.
To simplify the presentation we shall restrict to
large $N_c$ and set $N_f = x N_c$, and we shall take $k\leq5$ and
$3N_c/(k+1) < N_f < 2N_c$ to avoid complications of accidental
symmetry. 
The upper limit is the na${\ddot i}$ve asymptotic freedom condition.
Many more cases were actually studied with results in the
same pattern we report here.

In the large $N_c$, $N_f$ region the value of $a_{IR}$ 
in (\ref{acKS}) for  the
case $W= {\rm Tr}~ X^{k+1}$ is 
\begin{equation}
  \label{eq:5.1}
  a(k+1) = \frac{9N_c^2}{32} \left\{ 1 + \frac{2}{3(k+1)} - (1-
    \frac{2}{k+1})^3 - \frac{16}{(k+1)^3 x^2} \right\}
\end{equation}
which is positive in the region indicated above. 
The $S$-current method by which this value is computed
implicitly assumes that there is a free ultraviolet fixed point
and that the $S_{\mu}$ current is well defined along the RG flow.
If we make this assumption then the $a$-theorem is satisfied
for the flow from this fixed point 
since $r<5/3$ for both adjoint and fundamentals.

We can also test the $a$-theorem for flows which interpolate
between non-trivial fixed points in the Kutasov-Schwimmer series.
Indeed, evidence was given in \cite{kut,kss} that
in the perturbed theory with $W={\rm Tr}
X^{k+1} + {\rm Tr} X^k$, there are flows
from the $(k+1)$-fixed point theory in the UV (where 
${\rm Tr}X^k$ is
an irrelevant operator) to the $k$-fixed point theory in the IR
(where ${\rm Tr}X^{k+1}$ is irrelevant). 
Therefore the differences $a
(k+1) - a(k)$ provide further tests of the theorem in the new situation
of interacting critical theories at {\it both} ends of the
flow. 
The differences and their signs are as follows:
\begin{eqnarray}
  a(3) - a(2)= \frac{9N^2_c}{32} \left[ -.148 +
    \frac{1.407}{x^2} \right] & >0 ,& ~~~\frac{3}{2} < x <2;\nonumber\\
 a(4) - a(3) = \frac{9N^2_c}{32} \left[-.143 +\frac{.342}{x^2}
 \right] & <0 , &~~~1.546 <x <2 
\nonumber\\
&>0,&~~~1<x<1.546;\nonumber\\
a(5) - a(4) = \frac{9N^2_c}{32} \left[-.125 + \frac{.122}{x^2}
\right]  &<0 , &~~~ .988 < x <2\nonumber\\
&>0,& ~~~ .75 < x<.988;\nonumber\\
a(6) - a(5) = \frac{9N^2_c}{32} \left[-.102 + \frac{.054}{x^2}
\right] & <0 ,& ~~~ .728 <x<2\nonumber\\
&>0,& ~~~ .6<x<.728.
\end{eqnarray}
We thus observe additional violations of the $a$-theorem, which
occur in the 3 non-renormalizable cases above for $x$ in the upper
part of its allowed range. 
We will discuss this below, but let us
digress briefly to discuss a special property of the 
$W={\rm Tr} X^4$ theory,
which will strengthen our inference that failure of the
$a$-theorem is due to non-renormalizability.

We consider a theory whose field content is that of 
the Kutasov-Schwimmer
model with an extra chiral superfield $Y^{\alpha}_{\beta}$ in the
reducible $adj\oplus {\bf 1}$ representation of the gauge group. The
superpotential is $W = - {\rm Tr} Y^2 + 2 {\rm Tr} (Y X^2)$. 
The field $Y$ is
massive and may be integrated out to give $W_{eff} = {\rm Tr}
X^4$. 
Thus the new theory is equivalent to the $W= {\rm Tr} X^4$ 
Kutasov-Schwimmer
theory in the infrared, and is renormalizable, asymptotically
free and without  accidental symmetry in the reduced range 
$3N_c/4<N_f <N_c$. 
In the presence of the new chiral superfield $Y$ the value of $a_{UV}$ changes 
so that for the flow from the ultraviolet free fixed point to
the infrared we have
$$\Delta~a={7\over {768}} - {{N_fN_c}\over {24}} + 
  {{49\,{N_c^2}}\over {768}} - {{9\,{N_c^4}}\over {128\,{N_f^2}}}>0.$$
The computation above for $a(4) - a(3)$ was
valid only for $x>1$ because we did not include accidental
contributions. 
However we can now add the previously computed
contribution to $a(3)$ namely $\Delta a(3) = N^2_c (1-x)^2
(4-x)/6x$ (which should be multiplied by a step function
$\theta (1-x)$). 
The new result for the flow of $a$, namely $a(4)
-a(3) -\Delta a(3)$ is now valid for $.75 N_c < N_f <N_c$ and is
positive in this range. 
So the observed violation above occurs
only in the non-renormalizable region.

We must consider the question whether one can expect the
$a$-theorem to hold for non-renormalizable theories. 
In two
dimensions, Zamolodchikov assumed Wilsonian renormalizability in
his proof of the two-dimensional $c$-theorem. The structure of
the theory above some large cutoff $\Lambda$ was not relevant to his
demonstration that the $c$-function $C(g(\mu))$ is monotonically
decreasing toward the infrared below this scale. In the approach
of Cappelli, Friedan and Latorre \cite{Cappelli}
the ultraviolet central charge
$c_{UV}$ is expressed as an integral over a Lehmann weight function,
and the integral diverges in a (power-counting) 
 non-renormalizable
two-dimensional theory. The well known Cardy sum rule
$
c_{UV}-c_{IR}\sim \int d^2 x~~ x^2\langle \Theta (x) \Theta (0)\rangle
$
also diverges. It is entirely possible that in future work an
$A$-function
can be identified and monotonicity proven without
assumptions concerning the ultraviolet behavior. However, at present
we have theoretical control of the Euler anomaly coefficient only
at fixed points, and one must expect that this control is lost in the
ultraviolet limit of
a non-renormalizable theory. One possible technical reason is a problem
with the $S$-current method we have used. 
The $S$-current can be viewed as
the solution of the operator mixing problem for the current $R^{\mu}$.
In a renormalizable theory it can mix only with a flavor singlet
combination of Konishi currents, but in a non-renormalizable theory
there are an infinite number of possibilities.

\section{Theories with additional global U(1) symmetries}

In theories with anomaly-free global U$(1)_F$ symmetries the $R$-symmetry
is not unique and we {\em a priori} do not know which $R$-symmetry
participates in the superconformal algebra of the infrared theory.
As a result we cannot  determine
$a_{IR}$, $b_{IR}$ and $c_{IR}$ by the procedure described above.  
For simplicity we assume that there is a single U$(1)_F$ symmetry.
In this situation the anomaly free $R$-current is not unique, and there is a
one parameter ambiguity in the choice of constants $\gamma_i^*$
in the anomaly-free $S_0^\mu$ and 
$S^\mu$ currents of (\ref{2.18}) and (\ref{2.20}). 
We choose 
any member of this one-parameter
family as a particular $R$-symmetry with current $\overline{S}^{\mu}$. 
This corresponds to
a particular assignment of $R$-charges 
$\overline{r}_i = (2 + \gamma_i^*)/3$
for chiral
superfields $\Phi_i^\alpha$, each of which has a unique flavor charge $q_i$.
The most general $R$-current is then 
$S^\mu = \overline{S}^\mu - v J^\mu$ 
where $v$ is a real
parameter and $J^\mu$ is the flavor current, 
and the $R$-charges for this 
current
are $r_i(v)$ =  $\overline{r}_i -vq_i$. 
For one particular value of $v$ this
$S$-current is in the same multiplet 
as the stress tensor at the IR fixed  
point, but it is usually possible to determine 
$v$ only near the weakly coupled
end of the conformal window, where the RG flow is perturbative.
    
We can compute the anomaly coefficients $a_{IR}(v)$, $b_{IR}(v)$, 
$c_{IR}(v)$ as 
functions of $v$ from (\ref{abcIR}) and use the various positivity 
conditions to
constrain the value of $v$. 
A weak check 
of the $a$-theorem and conformality is then  
provided by
the constraint that there exist a region in $v$ for which all of  
the positivity
conditions are satisfied.
Conversely, these positivity conditions
constrain the scaling dimensions of operators at the fixed point.
Furthermore, the physically allowed value of $v$ is
restricted by the assumption that all chiral composite fields have $r(v)>2/3$
so that unitarity is satisfied without accidental symmetry.  

We now illustrate this procedure for the Sp($2N_c$) gauge theories with $2N_f$
fundamentals and one  two-index
symmetric tensor, previously studied in \cite{luty}, where evidence for a
non-Abelian Coulomb phase was given in the conformal window $0<N_f<2N_c+2$.  
The charges of the fields
under the global symmetries are given below, with a simple
choice for the anomaly-free $\overline{S}$-symmetry,
$$
\begin{array}{c||c||c|c|c}
& Sp(2N_c) & SU(2N_f) & U(1)_F & U(1)_{\tilde{S}} \\ \hline
S & \Ysymm & 1 & -1 & 0 \\
Q & \Yfund & \Yfund & \frac{N_c+1}{N_f} & 1
\end{array}
$$
As discussed above the value
of $v$ is constrained by unitarity. 
For this model $Q^2$ and
$S^2$ must have
scaling dimension greater than one, or $R$-charge greater than 2/3. 
 This
requires $v$ to lie in the range 
\begin{equation}
\frac{1}{3}< v < \frac{2N_f}{3(N_c+1)}, \label{eq:v} 
\end{equation}
and also determines the lower limit on $N_f$ in 
\begin{equation}
\frac{N_c+1}{2}<N_f<2(N_c+1), \label{eq:nf} 
\end{equation}
where the upper bound is from asymptotic freedom.  
Equations 
(\ref{eq:v}) and (\ref{eq:nf})  determine
the triangular ``physical region'' of the two parameters $N_f$ and $v$.
It is actually expected \cite{luty} that $v$ exits from the triangular
physical region below some value of $N_f$. 
In this case an accidental
symmetry is required, and our analysis is valid only above this value of
$N_f$.
In the $v-N_f$ plane we  plot the curves $c_{IR}(v,N_f)=0$ and 
$a_{IR}(v,N_f)=0$ 
for various values of $N_c$. 
The results, shown in Figs. 1-3, indicate that 
positivity $c_{IR}>0$ and $a_{IR}>0$ holds in 
the entire physical region. 
Further,
the flow $a_{UV}-a_{IR}$ and the value of $b_{IR}$ 
for both SU$(N_f)$ and 
U$(1)_F$
central charges is positive in the entire region shown. 
Thus there is no
constraint on the parameter $v$ from any of the positivity conditions
studied.

\vspace{.2in}

\vbox{

\hskip -2in\hbox{\hbox{
\let\picnaturalsize=N
\def\picsize{3in}
\def\picfilename{josh.eps}
\ifx\nopictures Y\else{\ifx\epsfloaded Y\else\fi
\global\let\epsfloaded=Y
\centerline{\ifx\picnaturalsize N\epsfxsize \picsize\fi
\epsfbox{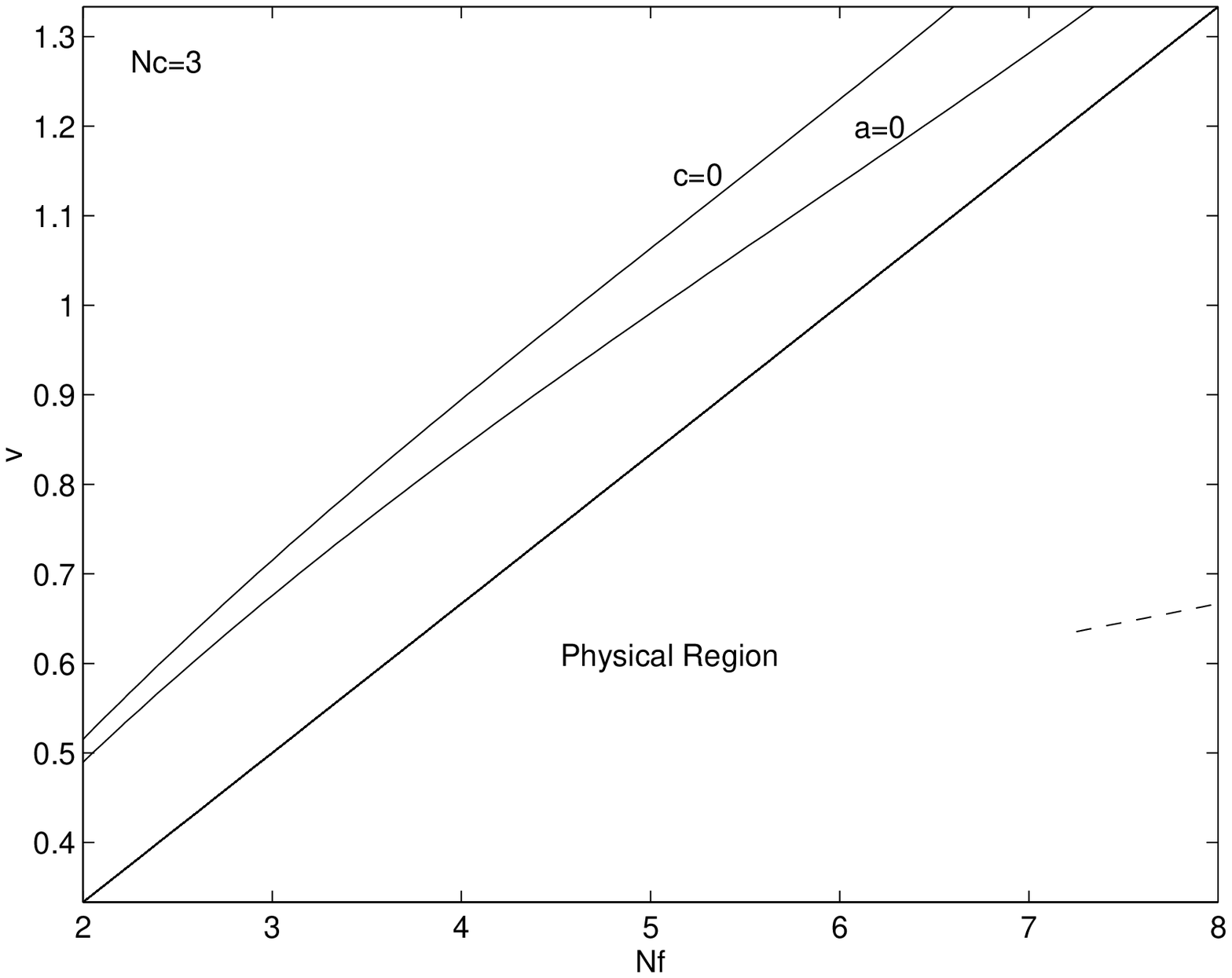}}}\fi
}
\hskip -3.5in\hbox{
\let\picnaturalsize=N
\def\picsize{3in}
\def\picfilename{josh2.eps}
\ifx\nopictures Y\else{\ifx\epsfloaded Y\else\fi
\global\let\epsfloaded=Y
\centerline{\ifx\picnaturalsize N\epsfxsize \picsize\fi
\epsfbox{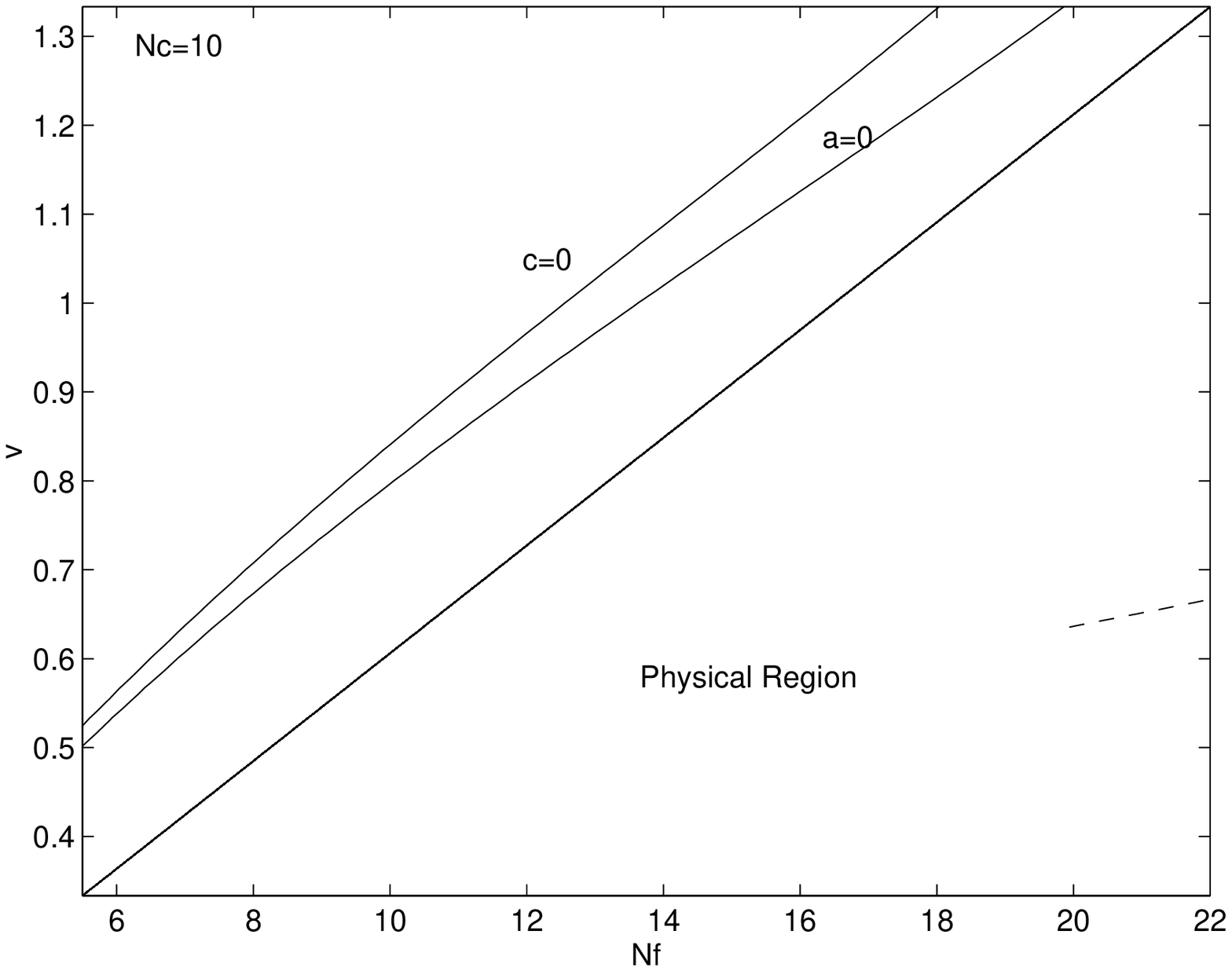}}}\fi
}}

\hskip -.5in\hbox{
\let\picnaturalsize=N
\def\picsize{3in}
\def\picfilename{josh3.eps}
\ifx\nopictures Y\else{\ifx\epsfloaded Y\else\fi
\global\let\epsfloaded=Y
\centerline{\ifx\picnaturalsize N\epsfxsize \picsize\fi
\epsfbox{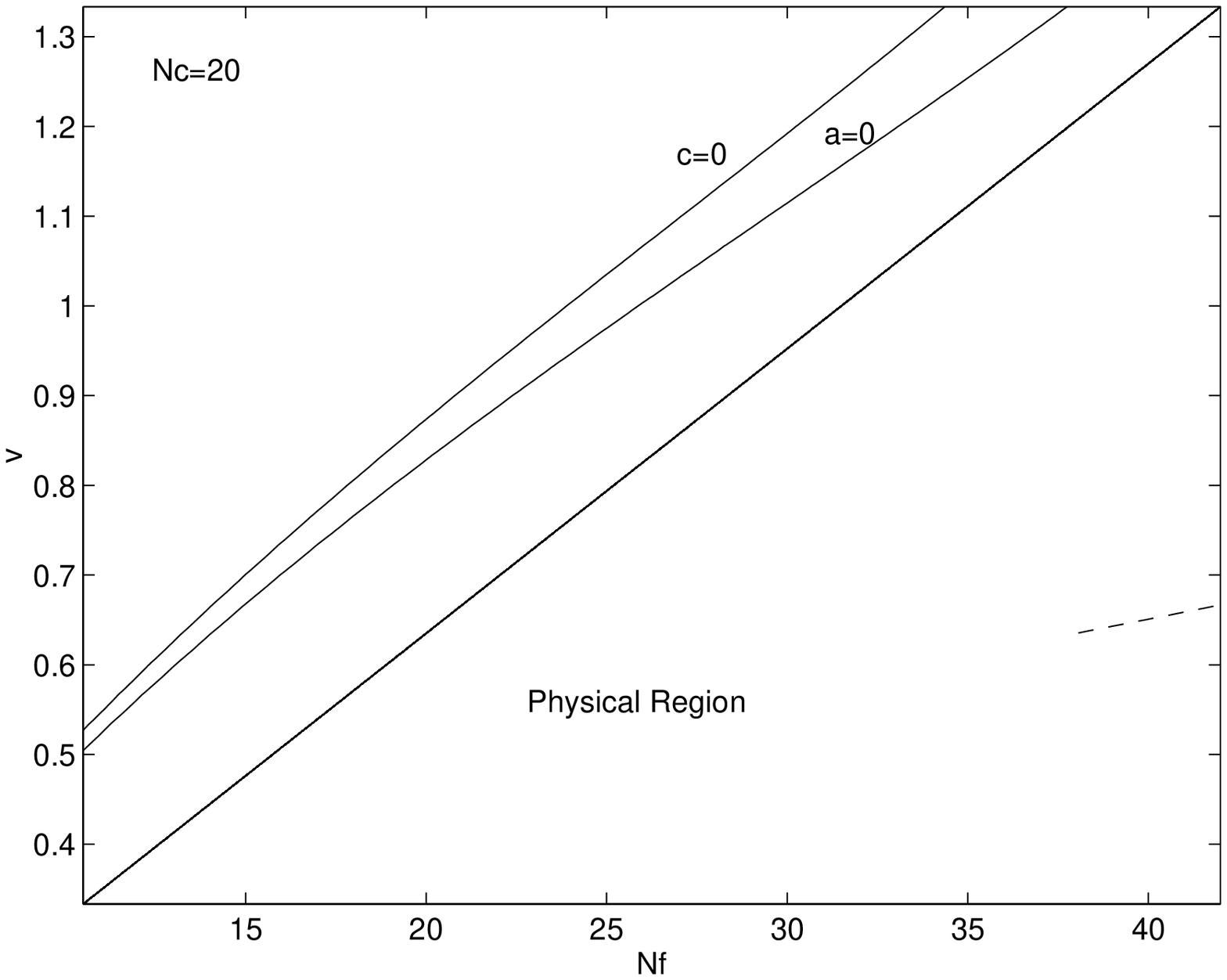}}}\fi
}

\centerline{Fig. 2. Positivity conditions are satisfied below 
the $c=0$ and $a=0$ curves,} 
\hspace{-.22in}\centerline{which includes the entire physical region. 
The short dashed line is the weak}

\hspace{-.2in}\centerline{coupling limit of $v$ from (\ref{7.46}).
Results are shown for various $N_c$. The flow }

\hspace{-.8in}\centerline{$a_{UV}-a_{IR}$ and $b_{IR}$ 
are positive everywhere in the graphs.}
\vspace{.2in}

}
\vspace{.2in}

Near the edge of the conformal window, {\em i.e.} 
near the upper bound for
$N_f$, we can determine the scaling dimensions of operators  
perturbatively,
hence determining the correct $R$-current order by order in the  
gauge coupling
at the fixed point, $\alpha_*$.  
The anomalous dimensions for the  
operators
$Q^2$ and $S^2$ are, near the point $N_f/(N_c+1)\sim 2$  
\cite{Banks-Zaks,luty},
 \begin{eqnarray}
\gamma_{S^2}&=&-\frac{\alpha_*}{\pi}(N_c+1)+O(\alpha_*^2) \nonumber \\
\gamma_{Q^2}&=&-\frac{\alpha_*}{2\pi}\left(N_c+\frac{1}{2}\right)+O(\alpha_*^2).
\end{eqnarray}
Defining $\varepsilon=2-N_f/(N_c+1)$, vanishing of the beta function,
$\beta\propto 4(N_c+1)-2N_f+2(N_c+1)\gamma_{S^2}+2N_f\gamma_{Q^2}$,  
to order
$\varepsilon$ determines the gauge coupling and anomalous dimensions,
\begin{equation}
-\gamma_{S^2}=\frac{\alpha_*}{\pi}(N_c+1)=\frac{\varepsilon}{2}+O(\varepsilon^2).
\end{equation}
Since the scaling dimensions are proportional to the $R$-charges,  
this fixes
$v$ to be
 \begin{equation}
v=\frac{2}{3}\left(1+\frac{\gamma_{S^2}}{2}\right)=\frac{2}{3}-
\frac{\varepsilon}{6}.
\label{7.46} \end{equation}

At the point $N_f=2(N_c+1)$, $a_{UV} -a_{IR}=0$.  This point is a local
minimum as a function of $N_f$ and $v$, so the flow is necessarily  
positive
as $v$ moves away from the free field value. 
 In fact, the perturbative
analysis is certain to preserve positivity since $b_{IR}$, $c_{IR}$ and
$a_{IR}$ are large and positive near the free point.

\section{Review of Results}

Let us summarize the conclusions of this paper. 
There are rigorous positivity
constraints on the flavor current and $Weyl^2$ trace anomaly coefficients in any 
renormalizable four-dimensional theory which flows from 
a conformal theory in the UV to another
in the IR. 
These constraints arise because the fixed-point values of the
anomaly coefficients coincide with central charges of the conformal algebra
at the fixed point, and the central charges must be positive by unitarity.
This part of the argument was first presented in \cite{gof97}. 
There are additional
conjectured positivity conditions \cite{l-a} on the Euler anomaly coefficient  
$a(g(\mu)) $ and
on its flow \cite{cardy} from the UV to the IR. 
In particular the
only viable candidate for a universal $c$-theorem in four dimensions seems
to be the inequality $a_{UV}-a_{IR} >0.$ 
There is no proof of this result, so it
is important to test it in models where both the UV and IR behavior are known.
It is fortunate that many such models are now known from the study of 
$N=1$
Seiberg duality. 
Because of asymptotic freedom the UV values of the anomaly 
coefficients can be simply obtained from lowest order 1-loop graphs, but
the IR values are more difficult because the coupling is strong at long
distance. It was first shown in \cite{bof97} that the IR values can be easily 
computed from the U$(1)_RFF$, U$(1)_R$, and U$(1)_R^3$ 
anomalies which are usually
calculated to establish the IR equivalence of the electric and magnetic duals.
This is possible because of the close relation between the trace anomaly and 
the anomalous divergence of the U$(1)_R$ current in global and local 
supersymmetry. 
Results \cite{bof97} of tests of the positivity conditions in the
SU$(N_c)$ series of SUSY gauge theories showed that all conditions were satisfied
throughout the conformal window, 
and that other possible $c$-theorem
candidates could be ruled out.

The major purpose of the present paper was to test the positivity
constraints in many more models.
For this purpose we developed general
formulae (\ref{abcIR}) for the infrared anomaly coefficients in terms 
of the anomaly
free $R$-charges. 
In models where the non-anomalous
$R$-charges are unique, a precise test of the positivity 
conditions can be carried
out with little difficulty, and this has been done for 
the rigorous  conditions
$b_{IR}>0$ and $c_{IR}>0$ for flavor and $Weyl^2$ anomalies, 
as well as the $a$-theorem
itself and the associated condition $a_{IR}>0.$
In many cases positivity can be
established from rather weak sufficient conditions, 
but a closer analysis is
required for models with accidental symmetry and for 
flows between interacting
fixed points generated by a relevant perturbation or 
Higgs deformations of
the UV fixed point theory. 
All conditions are satisfied in the large number of
renormalizable theories we have studied, but there are 
counterexamples for
interpolating flows in non-renormalizable theories 
where $a_{UV}-a_{IR}$ can
have either sign. 
There is 
considerably less theoretical control in non-renormalizable cases and, even
in two dimensions, tests of the $c$-theorem which involve the ultraviolet limit
of a power-counting non-renormalizable theory seem to be problematic. 
Provisionally, then, we believe that the cases of negative flows in 
non-renormalizable should not be viewed as ruling out a universal $a$-theorem.

The assignment of $R$-charges in theories conjectured to 
be in the non-Abelian
Coulomb phase is important for the understanding of infrared dynamics because
the N=1 superconformal algebra necessarily includes the generator of U$(1)_R$  
transformations. 
This assignment is not guaranteed to satisfy the rigorous
positivity conditions, and the fact that these are satisfied is a broad
consistency check of $N=1$ duality. 
The fact that $a_{IR}>0$ and 
$a_{UV}-a_{IR} >0$ in all renoromalizable models is very strong 
evidence that there
is a universal $a$-theorem, and that the RG flow is irreversible 
in four-dimensional supersymmetric theories, and perhaps more. 
We hope that this
empirical result might stimulate a successful theoretical proof.

It is worth noting that the present approach is not
immediately applicable to some  superconformal models
with $N=2$  
\cite{douglas,plesser,minahan,gaga} and $N=1$ \cite{kachru}.
It would be interesting to extend the present method to these cases.
Note that an approach to the computation of the flavor $b_{IR}$
in the $N=2$ theories has been
recently suggested in \cite{ganor}.

\section{Acknowledgements}

We are grateful to Michael Bershadsky, Asad Naqvi, Csaba Csaki, Witold Skiba,
Matthew Strassler, Cumrun Vafa, and Fabian Waleffe for useful
discussions.
We also thank 
The research of D.A. was
partially supported by EEC grants CHRX-CT93-0340 and TMR-516055. 
The research of J.E. was supported by DOE cooperative research agreement
DE-FC02-94ER40818.
The
research of D.Z.F. was supported by NSF grant PHY-97-22072.
The research of
A.A.J. was
supported in part by the Packard Foundation and by NSF grant
PHY-92-18167.
\vspace{0.2in}

\appendix

\section{Appendix: Tests of a possible $b$-theorem}
\renewcommand{\theequation}{A.\arabic{equation}}
\setcounter{equation}{0}

We present here tests of the inequality $b_{UV}-b_{IR}>0$ for the flow of
flavor current central charges in the situations i-iv for which
previous tests of the $a$-theorem were discussed in Sec. 2.

\begin{romanlist}
\item Let us assume (as was done in \cite{cardy}) that SU$(N_c$) QCD is
realized in a confined phase with chiral symmetry breaking, so the massless 
spectrum consists of $N_f^2-1$ Goldstone bosons which decouple in the
long distance limit.  For the baryon number current one clearly has
$b_{UV}-b_{IR}>0$ since there are no massless baryons.  For a current of the
vectorial SU($N_f$) flavor group, on the other hand, we find $b_{UV}\propto
4N_c$ and $b_{IR}\propto N_f$ with a common constant of proportionality.
Thus $b_{UV}-b_{IR}$ changes sign within the region of asymptotic freedom.
Of course this could just mean that the conjectured Goldstone realization
fails for $4N_c<N_f<11N_c/2$.

\item To investigate the $b$-theorem for large $N_c,\,N_f$ we can
make use of the well known QED $\beta$-function.  
Up to two-loop order it
is given by $\beta_{QED}(\alpha)=
\frac{2\alpha^2}{3\pi}+\frac{\alpha^3}{2\pi^2}$.  
The graphs for the flavor current correlator in QCD are obtained from
the identical QED graphs (see Fig. 3) by replacing the U(1) coupling
by the SU($N_f$) flavor matrix $T^A/2$ at each external vertex and by
the gauge coupling matrix $gt^a/2$, where $t^a$ is an SU($N_c$) color
matrix at each internal vertex.  
The point is that these replacements preserve
the relative positive sign between the one and two-loop contributions. 

\vskip .1in

\vbox{
\let\picnaturalsize=N
\def\picsize{5in}
\def\picfilename{QED.eps}
\ifx\nopictures Y\else{\ifx\epsfloaded Y\else\fi
\global\let\epsfloaded=Y
\centerline{\ifx\picnaturalsize N\epsfxsize \picsize\fi
\epsfbox{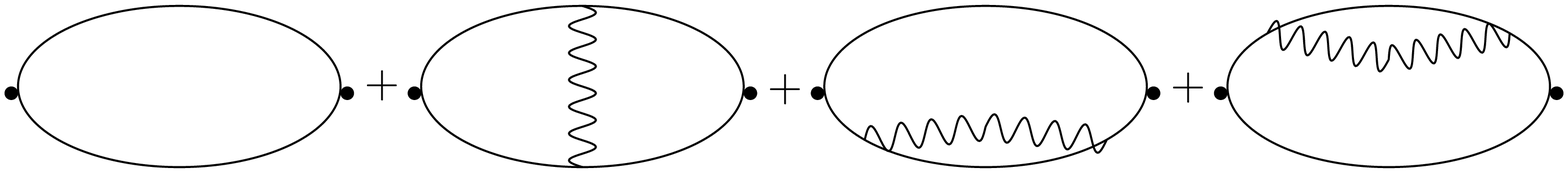}}}\fi
\vskip .1in

\centerline{Fig 3. The graphs for the flavor current correlator.}
}
\vskip .1in

The
current correlator then takes the form \begin{equation}
\langle J_\mu^A(x)J_\nu^A(x)\rangle\sim(\Box\delta_{\mu\nu}-\partial_\mu
\partial_\nu)\frac{{\rm Tr}\,(T^A)^2}{x^4}[N_c+\rho g^{\ast\,2}]
\end{equation}
where $\rho$ is a positive constant and the 
fixed point value of the coupling 
is $\frac{g^{\ast\,2}}{4\pi}=\frac{22N_c-4N_f}{75N_c^2}$.  
The same is true for the correlator of
baryon number currents.  
Thus $b_{UV}-b_{IR}\sim[N_c-(N_c+\rho g^{\ast\,2})]<0.$

\item One may also test a possible $b$-theorem in the free magnetic phase
of SU$(N_c$) SUSY QCD as follows.
In the ultraviolet we compute $b_{UV}$ from the free field $\langle R^\mu J^\nu
J^\rho\rangle$ correlator in the electric theory.  The infrared value $b_{IR}$
is obtained from a similar free field computation in the magnetic theory.
The difference is 
\begin{equation}
b_{UV}-b_{IR}=-\frac{1}{3}\left[\frac{2N_c^2N_f}{N_f-N_c}-2N_fN_c\right]
=\frac{2N_fN_c}{N_f-N_c}\,[N_f-2N_c],
\end{equation}
which is negative in the entire free magnetic region.  Hence the $b$-theorem
fails again. 

\item In the entire conformal window $3N_c/2<N_f<3N_c$ of SU($N_c$) 
supersymmetric QCD, it is known \cite{bof97} that $b_{UV}-b_{IR}<0$
in both electric and magnetic theories for the baryon number central charge.
We present here a more general computation for an electric type theory
with $N_f$ copies of $(R\oplus\overline{R})$, and we include a mass
deformation, making $n$ flavors massive.  
For a current of the low
energy SU$(N_f-n)$ flavor group, we have, using ${\rm Tr}\,(T^A)^2=1/2$,
the central charges $b_{UV}={\rm dim}\,R$ at the free UV point, and
$$b_n=3\,{\rm dim}\,R~\frac{T(G)}{2(N_f-n)~T(R)}$$ 
for the interacting fixed
point theory with $N_f-n$ massless flavors.  
One can then see that 
asymptotic freedom implies $b_{UV}-b_n<0$ so the $b$-theorem fails for
a flow from the free UV fixed point to any of the IR fixed point theories.
Furthermore $b_{n_1}-b_{n_2}<0$ if $n_1<n_2$, so the flow between
any pair of fixed point theories in which the number of massless quarks
decreases also violates a $b$-theorem.  
At this point one might think that
an anti-$b$-theorem holds in supersymmetric theories. 
However this is not the
case for Higgs deformations.  
To see this we consider the basic Higgs 
deformation of the SU$(N_c)$, SU$(N_f)$ 
theory, leading to the SU$(N_c-1)$, 
SU$(N_f-1)$ IR theory plus $2(N_f-1)$ decoupled Goldstone fields.  For an
SU$(N_f-1)$ flavor current we have $b_{UV}=N_c$ at the free UV point, while
$b^\ast_{IR}=3(N_c-1)^2/(N_f-1)+1$ in the Higgsed low energy theory.
The contribution $+1$ comes from Goldstone fields. One sees quite easily 
that $b_{UV}-b^\ast_{IR}$ can have either sign in the conformal window,
and the same is true for the flow from the SU$(N_c)$, SU$(N_f)$ fixed
point to that of the Higgsed theory.
\end{romanlist}

The conclusion of this analysis is that the flow of flavor central charges
does not have a recognizable universal property.

\end{document}